\documentclass[twocolumn,pre,aps,superscriptaddress,nofootinbib]{revtex4-2}
\usepackage[utf8]{inputenc}

\usepackage{graphicx}% Include figure files
\usepackage{dcolumn}% Align table columns on decimal point
\usepackage{gensymb}%For the degree symbol
\usepackage{svg}
\usepackage{xcolor}
\usepackage{amsmath,amssymb,dsfont,amstext,amsfonts}
\usepackage[colorlinks=true]{hyperref}  
\usepackage{bigints}

\usepackage[capitalize]{cleveref}   % Améliore les ref + ref cliquables
\usepackage{bm}% bold math
\usepackage{physics}% Include figure files

\usepackage{appendix}
\hypersetup{
    bookmarks=true,         % show bookmarks bar?
    unicode=false,          % non-Latin characters 
    pdftoolbar=true,        % show Acrobat
    pdfmenubar=true,        % show Acrobat 
    pdffitwindow=false,     % window fit to page when opened
    pdfstartview={FitH},    % fits the width of the page to the window
    pdfsubject={},   % subject of the document
    pdfcreator={},   % creator of the document
    pdfproducer={}, % producer of the document
    pdfkeywords={} {} {}, % list of keywords
    pdfnewwindow=true,      % links in new window
    colorlinks=true,       % false: boxed links; true: colored links
    linkcolor=blue, %red,          % color of internal links (change box color with linkbordercolor)
    citecolor=blue,        % color of links to bibliography
    filecolor=magenta,      % color of file links
    urlcolor=blue,           % color of external links
} 

\begin{document}
\title{Quasiperiodicity Protects Quantized Transport in Disordered Systems Without Gaps}
\author{Emmanuel Gottlob}
\email{emg69@cam.ac.uk}
 \affiliation{Cavendish Laboratory, University of Cambridge, JJ Thomson Avenue, Cambridge CB3 0HE, United Kingdom}

\author{Dan S. Borgnia}
\email{dborgnia@berkeley.edu}
\affiliation{Department of Physics, University of California, Berkeley, California 94720, USA}
\author{Robert-Jan Slager}
\email{rjs269@cam.ac.uk}
\affiliation{TCM Group, Cavendish Laboratory, Department of Physics, JJ Thomson Avenue, Cambridge CB3 0HE, United Kingdom}
\author{Ulrich Schneider}
\email{uws20@cam.ac.uk}
 \affiliation{Cavendish Laboratory, University of Cambridge, JJ Thomson Avenue, Cambridge CB3 0HE, United Kingdom}

\date{\today}

\begin{abstract}
The robustness of topological properties, such as  quantized currents, generally depends on the existence of gaps surrounding the relevant energy levels or on symmetry-forbidden transitions.
Here, we observe quantized currents that survive the addition of bounded local disorder beyond the closing of the relevant instantaneous energy gaps in a driven Aubry-Andr\'e-Harper chain, a prototypical model of quasiperiodic systems. We explain the robustness using a local picture in \textit{configuration-space} based on Landau-Zener transitions, which rests on the Anderson localisation of the eigenstates. Moreover, we propose a protocol, directly realizable in for instance cold atoms or photonic experiments, that leverages this stability to prepare topological many-body states with high Chern numbers and opens new experimental avenues for the study of both the integer and fractional quantum Hall effects.
\end{abstract}
\maketitle

A key feature of topological phases is the robustness to local perturbations that leads to  quantized physical observables characterized by a bulk topological invariant, a famous example being the integer quantum Hall effect~\cite{ThoulessQHE, halperin1982quantized}.
Topological phases can also be realised in dynamical systems, where time is interpreted as an additional \textit{virtual} dimension. The canonical example is Thouless pumping \cite{thoulessQuantizationParticleTransport1983,citroThoulessPumpingTopology2023b}, where a slow time-periodic modulation of a typically 1D Hamiltonian generates quantized charge transport indexed by \textit{virtual} 2D Chern numbers \cite{ kraus2012topological,prodan2015}. Experimentally, Thouless pumps have been realized within various platforms including photonic waveguides~\cite{Ozawa2019} and ultracold atoms \cite{ lohseThoulessQuantumPump2016a,nakajimaTopologicalThoulessPumping2016, Walter2023,  viebahntopopump}.
They are typically considered in the context of periodic systems, where the pumped charge is only quantized when the pumping time is an exact multiple of the pump period. In contrast to periodic systems, however, Thouless pumping in \textit{quasiperiodic} systems~\citep{kraus2012topological, marraTopologicallyQuantizedCurrent2020, moustaj2024anomalouspolarizationonedimensionalaperiodic} can exhibit constant quantized pumped charge and currents at all times during the pump cycle~\citep{marraTopologicallyQuantizedCurrent2020}. 

The stability of quantized Thouless pumping to disorder remains a topic of active research \cite{citroThoulessPumpingTopology2023b,grabarits2023floquetanderson,vuina2024absence}.
It has long been proven that the adiabatic theorem does not require a spectral gap~\cite{avron1999adiabatic}, even in open quantum systems~\cite{avron2012adiabatic}. For experimentally accessible observables such as the pumped charge, however, topological quantization typically remains stable to generic perturbations only up to the closing of spectral or mobility gaps  between delocalized eigenstates \cite{PhysRevB.23.5632} or the breaking of explicit symmetry protections~\cite{10.1063/1.3149495,PhysRevX.7.041048,Clas3,PhysRevX.11.041059}. In particular, it was shown that Thouless pumps in periodic systems break down under closing of the energy gaps by disorder~\cite{Wauters, nakajimaCompetitionInterplayTopology2021b,haywardEffectDisorderTopological2021a}. 

In this letter, we report on an experimentally accessible Thouless pump for quasiperiodic systems
characterized by persistent quantized pumping in the presence of large bounded local disorder, surviving the disorder-induced closure of spectral gaps. 
We discuss how the robustness can be readily probed within existing cold atoms experiments, and leveraged to prepare many-body states with arbitrary Chern number. 

To explain this counter-intuitive robustness, we develop a perturbative argument based on local Landau-Zener (avoided-crossing) transitions, and derive approximate  critical disorder strengths using a \textit{configuration-space} picture. The enhanced stability results from two special properties of 1D quasiperiodic Hamiltonians: First, its eigenstates can be Anderson localized~\cite{jitomirskaya1998anderson} even in the absence of additional disorder. Second, the spectra of these quasiperiodic operators form bounded Cantor sets.
Each spectrum contain a fractal hierarchy of gaps \cite{avila2009ten} whose integrated density of states (IDoS) is quantized by the gap labeling theorem \cite{bellissard92,drytenmartininoncritical,han2018dry}, giving rise to their unusual topological properties.

While our results generalize to other short-range, bounded, analytic quasiperiodic Hamiltonians, where eigenstates are Anderson localized and the gap labelling theorem holds~\cite{bellissard92,jitomirskaya1998anderson},
we focus on the prototypical Aubry-Andre-Harper (AAH) model.
This model is realized experimentally in, for instance, optical lattices~\cite{Roati2008, schreiber}, photonic systems~\cite{kraus2012topological}, or with cavity polaritons~\cite{Goblot2020}  and is known as the almost Mathieu operator in the mathematical physics literature
\begin{equation}\label{eq:AAH_clean}
	\hat{H}_{AAH}^{J,V} (\varphi) = \hat{V}(\varphi)-\hat{J}.
\end{equation}
It consists  of a nearest-neighbor hopping term\begin{align}
    \hat{J} = \sum_{i} J(\hat{a}^{\dagger}_{i+1}\hat{a}_i +\hat{a}^{\dagger}_{i}\hat{a}_{i+1})
\end{align}
and an onsite cosine potential 
\begin{align}\label{eq:V}
    \hat{V}(\varphi) = \sum_{i} 2V\cos(2\pi\beta i+\varphi)\hat{a}^{\dagger}_{i}\hat{a}_i,
\end{align}
where $i \in \mathbb{Z}$ denotes the lattice sites and $a_i$ ($a_i^\dagger)$ are the corresponding annihilation (creation) operators for single particles on site $i$. For irrational  $\beta\in \mathbb{R}\backslash\mathbb{Q}$, $\hat{V}(\varphi)$  breaks the translation invariance of $\hat{J}$ and the resulting eigenfunctions are Anderson localized for $\vert V/J\vert >1$~\cite{jitomirskaya1998anderson} and extended for $\vert V/J\vert <1$~\cite{jitomirskaya1999metal}, as a result of self-duality \citep{aubry1980annals} around the critical point $V/J=1$.  We restrict $\beta \in  (0,1)$ without loss of generality and set $\beta = \sqrt{2}/2$ in all figures. Thouless pumping is implemented by linearly increasing the phase $\varphi$ in time at a constant pumping rate $\dot{\varphi}>0$.

\begin{figure}
    \centering
    \includegraphics[width = \linewidth]{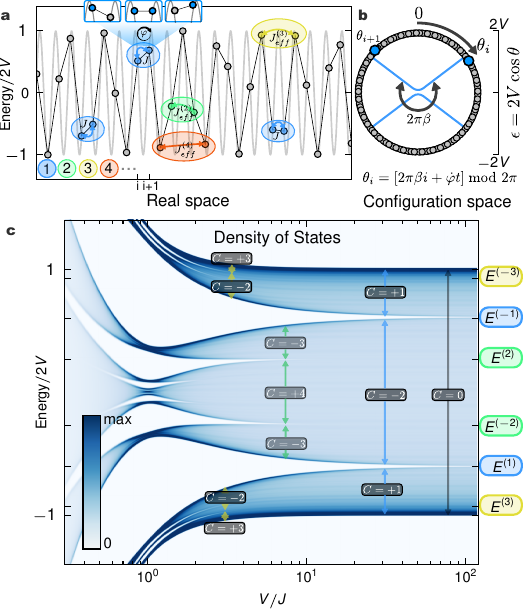}
    \caption{\textbf{Perturbative derivation of the AAH energy spectrum:} (\textbf{a}) AAH chain where grey line denotes the quasiperiodic modulation of onsite energies, and circled dots are lattice sites. Solid lines indicate sites connected by the tunneling $\hat{J}$. Resonances between $n$-th nearest neighbors (colored ellipses, blue for $n=1$) are mediated by an exponentially decreasing effective tunneling rate $J_{eff}^{(n)}\propto J^n/V^{n-1}$. (\textbf{b}) We map the  AAH chain onto a uniformly dense unit circle referred to as configuration space. In this representation, $n$-th order neighbors are separated by an angle $\theta_{i+n}-\theta_{i}=n 2\pi\beta$. Vertical direction denotes energy,  and the resonances occur at fixed angles $\theta^{(n)}$, where two $n$-th order neighbors have the same height.  Blue lines connect sites involved in $n=1$ resonances. (\textbf{c}) The resonances result in a dense set of infinitely many pairs of energy gaps labelled by the integer $n$, which for $V\gg J$ lie at energies $E^{(\pm n)}$ (we mark gaps up to $n=3$, but all gaps are open for all values of $V/J$). Sets of states surrounded by two selected gaps acquire a Chern number $C$ fixed by the labels of the surrounding gaps.
    \textit{Parameters:} $\beta = \sqrt{2}/2$.}

    \label{fig:fig1}
\end{figure}

\section{AAH: Energy gaps and configuration space} \label{section:configspace}

In periodic systems, the Brillouin zone offers a suitable base manifold on which to calculate topological invariants and band structures. To circumvent the lack of a Brillouin zone for quasiperiodic lattices, we construct a \textit{configuration-space} representation of the clean AAH chain, see \cref{fig:fig1}b. This representation offers a powerful language to discuss such quasiperiodic systems and enables a simple perturbative argument to derive the positions of the energy gaps, their labels, and exact expressions for their respective integrated densities of state.
It is based on the observation that the only degree of freedom that differentiates one lattice site from another is the local value of the phase $\theta$ of the quasiperiodic modulation in \cref{eq:V}. Therefore, labeling the sites with the local phase of the modulation offers a convenient description of the problem. In this configuration space representation, the position of site $i$ is mapped to the angle $\theta_i$ on the unit circle according to 
\begin{align}\label{eq:configspace}
    \theta_i = 2\pi \beta i + \varphi  \mod 2 \pi,
\end{align}
see \cref{fig:fig1}b.
Since $\beta$ is irrational, this map is ergodic
and the set of all possible $\theta_i$, $\lbrace\theta_{i}\rbrace_{i\in\mathbb{Z}}$, densely and uniformly populates the unit circle \citep{zorziElementaryProofEquidistribution2015}, implying that every site is identified by a unique value of $\theta$. 

The Hamiltonian in \cref{eq:AAH_clean} can then be re-expressed on the densely populated unit circle:
\begin{equation} \label{eq:AAHconfigspace}
    \hat{H}_{AAH}^{J,V}  = 2V\sum_{\theta} \cos \theta \,  \hat{a}^\dagger_\theta \hat{a}_\theta - J\sum_{\theta}(\hat{a}^\dagger_\theta \hat{a}_{\theta+2\pi\beta} + h.c.) \,,
\end{equation}
where $\hat{a}_\theta$ is the annihilation operator for a particle at coordinate $\theta$ in configuration space, and the second term reflects the fixed phase relation between real-space nearest-neighbors: $\theta_{i+1} - \theta_i = 2\pi \beta$. 
Importantly, sites that are close in configuration space can be arbitrarily far away in real space.
However, due to the continuity of the cosine function, they will not only have similar energies but will also be surrounded by similar local configurations of lattice sites in real space.

In the absence of tunneling ($J/V =0$), the spectrum 
\begin{align}
    \sigma(\hat{H}_{AAH}) = \bigcup_{i\in\mathbb{Z}}\left(2V\cos(\theta_{i})\right)
\end{align} is dense on the interval $\left[-2V, 2V \right]$ (see right-hand side of \cref{fig:fig1} c),
and every eigenstate is localized on a single lattice site. Turning on a small tunneling amplitude ($J/V \ll1$), resonant neighbors, i.e.\ nearest neighbors whose onsite energy difference is small $\delta E=2V\left|\cos{(2\pi\beta (i+1)+\varphi)}-\cos{(2\pi\beta i+\varphi)}\right|\ll J$ (marked in blue in \cref{fig:fig1} a) hybridize and form  (anti-) symmetric dimers, separated by the energy $\Delta E^{(1)} = 2J$. Nearly-resonant neighbors with similar onsite energies will be affected similarly.  

These resonances happen at fixed positions in configuration space, namely where neighboring sites at $\theta$ and  $\theta + 2\pi \beta$ have similar onsite energies, i.e.,
\begin{align}
    \cos{\theta^{(1)}} \approx \cos(\theta^{(1)}+ 2\pi \beta)\,,
\end{align}
and the involved sites are connected by  blue lines in \cref{fig:fig1} b. 
The resulting level repulsion opens a pair of spectral gaps  of size $\Delta E^{(1)} \approx 2J$ at energies $\pm 2V \cos{\pi \beta}$ that split the spectrum into three \textit{first-order bands} (indicated by blue arrows in \cref{fig:fig1} c).
The IDoS of each of these bands is equal to the angle spanned by the band in configuration space. For instance, the lowest and highest first-order bands therefore contain, for $\beta>1/2$, $(1-\beta)$ states each, while the middle band contains $\left(2\beta-1\right)$ states, where we normalise the total number of states to $1$.

This perturbative analysis can directly be extended by looking for pairs of resonant sites at increasing distances $n>1$. In $n$-th order, tunneling causes hybridisation of resonant $n$-th-nearest neighbors, mediated by an effective tunneling amplitude of order $J_{\textit{eff}}^{(n)} \propto J^n/V^{n-1}$. The corresponding resonance  
conditions are: 
\begin{align} \label{eq:resonance}
    \cos{\theta^{(n)}} \approx\cos(\theta^{(n)}+ n \, 2\pi \beta)\,,
\end{align}
which cause the opening of pairs of gaps of size $\Delta E^{(n)} \propto 2J^n/V^{n-1}$ at energies 
\begin{equation} \label{eq:gaps}
    E^{(\pm n)} = \begin{cases}
    \pm 2V \cos{(n \pi \beta)} \, \, \text{if } \, [n\beta] \, \text{even} \\
    \mp 2V \cos{(n \pi \beta)} \, \,  \text{if } \, [n\beta] \, \text{odd}
    \end{cases} \,.
\end{equation}
Here, $[n\beta]$ refers to the integer part of $n\beta$. \cref{eq:gaps} shows that the AAH spectrum contains an infinite hierarchy of pairs of gaps -- forming a dense set -- labelled by the integer Chern numbers $\pm n$, where the two branches of the equation ensure consistency between the gap labels and the direction of quantized currents (see \cref{app:gaplabels} for details).  
By the gap labeling theorem, every band, that is the set containing all states between two selected energy gaps, is indexed by a virtual Chern number \cite{prodan2015} fixed by the difference between the two surrounding gap labels (\cref{fig:fig1} c). This Chern number is referred to as ``virtual" in the sense that it is defined by treating $\varphi$ as a virtual dimension of the system. It arises from the non-commutativity of $x,\partial_{\varphi}$-translations in configuration space analogous to the non-commutativity of $x,y$-translations responsible for the integer quantum hall effect \cite{prodan2015,paper1,paper2}. 

While the above gap construction relies on the hierarchy of effective tunnelings $J_{\textit{eff}}^{(n+1)}\ll J_{\textit{eff}}^{(n)}$ to generate a hierarchy of spectral gaps, it was shown that the spectrum indeed forms a nowhere-dense Cantor set with constant IDoS  for all $V/J\neq 1$~\cite{avila2009ten,drytenmartininoncritical}, extending the results of the above analysis to any value of $V/J>1$ and by self-duality to $V/J<1$. 
\begin{figure}
    \centering
    \includegraphics[width = \linewidth]{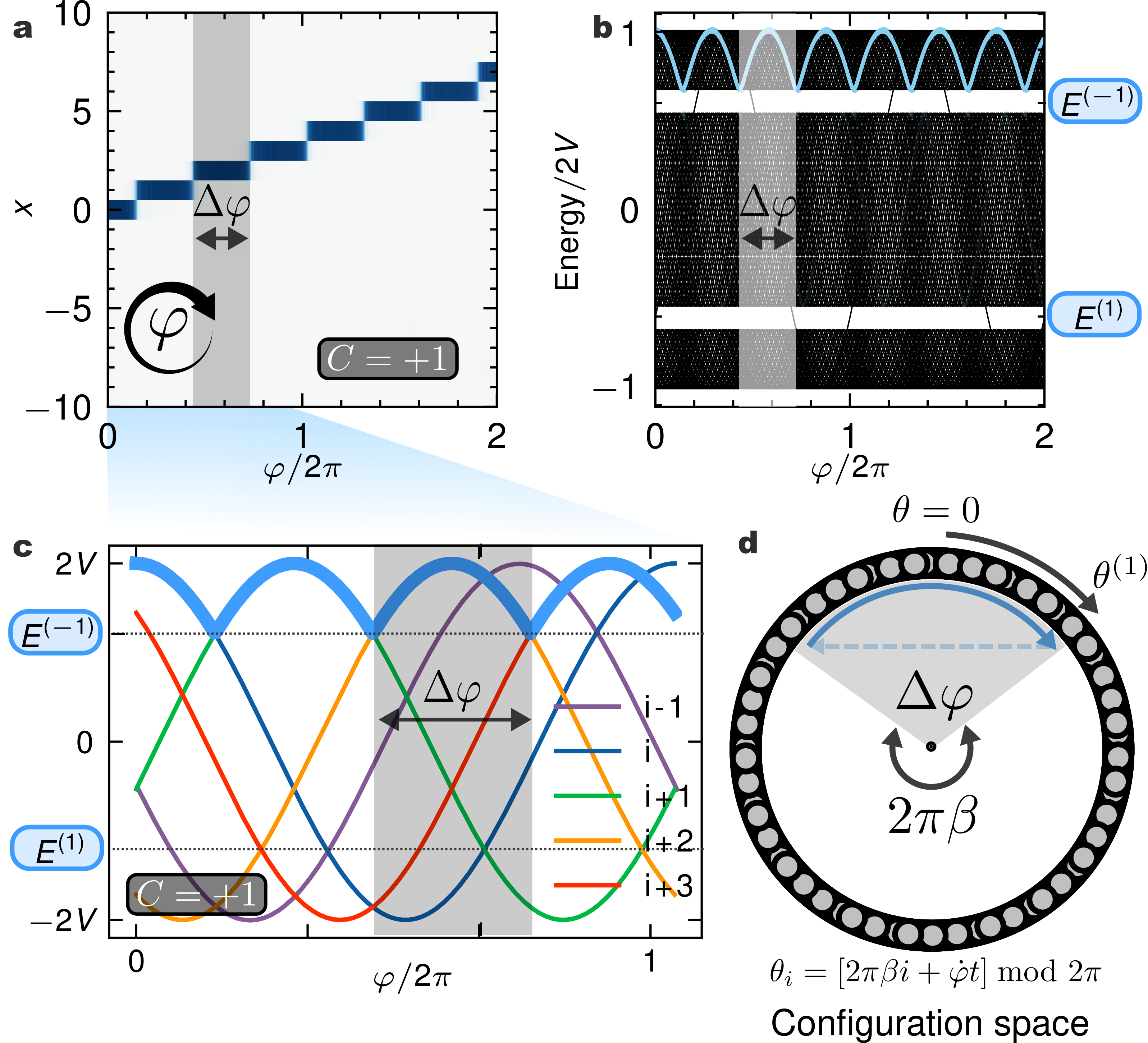}
    \caption{\textbf{Thouless pumping in the clean AAH chain:} (\textbf{a}) Thouless pumping of a single strongly localized state in the upper $C=+1$ band, shown in real space. (\textbf{b}) Due to adiabaticity with respect to the first-order gaps, the state remains confined within its energy band. Blue curve shows the projection of the wavefunction onto the instantaneous eigenstates shown in black. (\textbf{c})  The Thouless pump can be understood by noting that the particle undergoes an adiabatic Landau-Zener (light blue curve in \textbf{c}) every time it crosses in energy with a resonant $n$-th neighbor (here we show $n=1$). (\textbf{d}) In the configuration-space picture, pumping acts as a rotation (blue arrow); and particles in the top $C=+1$ band undergo one Landau-Zener transition (blue dashed arrow) for every $\Delta \varphi = 2\pi (1-\beta)$. The same argument can be extended to any of the energy bands. \textit{Parameters}: $\beta = \sqrt{2}/2$. $J = 1$, $V = 7.5$, $W/(2V) = 0$. $\dot{\varphi} = 5 \cdot 10^{-2}$. }
    \label{fig:fig_pumping}
\end{figure}

\section{Thouless Pumping in the AAH model}\label{qpthoulesssec}
Thouless pumping is a special case of adiabatic transport that occurs under a continuous, slow, and periodic modulation of the Hamiltonian. If the path in parameter space encloses a non-trivial topological charge, and the modulation is slow enough to not induce excitations out of the occupied subspace, the pumping gives rise to a quantized current proportional to the value of this charge~\cite{thoulessQuantizationParticleTransport1983}.

The Hamiltonian $\hat{H}^{J,V}_{AAH}(\varphi)$ is smooth and periodic in $\varphi$ and its spectrum forms a Cantor set  independent of $\varphi$. 
Since a Cantor set is nowhere dense, the spectrum can be separated into arbitrarily many ``bands" 
\begin{align}
    B^{(m,n)}\subset\sigma\left(\hat{H}^{J,V}_{AAH}(\varphi)\right),
\end{align} isolated by spectral gaps $\Delta E^{(m)},\Delta E^{(n)}$ at energies $E^{(m)},E^{(n)}$ independent of $\varphi\in[0,2\pi)$. This guarantees the existence of smooth contours $\gamma_{m,n}(\varphi)$ in the complex energy plane containing all of $B^{(m,n)}$ for all $\varphi$, which in turn guarantees adiabatic transport in the $\dot{\varphi}\rightarrow 0$ limit~\cite{Kato1950}. Thus, the 1D system evolving under $\hat{H}^{J,V}_{AAH}(\varphi(t))$ with $\varphi(t) = \dot{\varphi} t+\varphi_{0}$ and $(\hbar 2V\dot{\varphi})^{1/2}\ll\min(\Delta E^{(m)},\Delta E^{(n)})$ constitutes a Thouless pump with period $T=2\pi/\dot{\varphi}$ and results in a quantized current proportional to the band's Chern number. The current
\begin{equation}
    \hat{I}(t) = \frac{i}{L}J\sum_{i=1}^L (\hat{a}^{\dagger}_{i+1}\hat{a}_i -\hat{a}^{\dagger}_{i}\hat{a}_{i+1})
\end{equation}
 is the spatial average of the current operator over the system (of length $L$), and is equal to the derivative of the total pumped charge $I \equiv \partial_\varphi Q$ averaged over the entire system. In our numerics, we extract the pumped charge by fitting the evolution of the center-of-mass during pumping.  In contrast to periodic systems, where the current is time-dependent and only the pumped charge per pump cycle is quantized, in the quasiperiodic case the current itself is time independent and quantized  \citep{marraTopologicallyQuantizedCurrent2020} due to the invariance of the spectrum with respect to $\varphi$. 

For a given pump rate $\dot{\varphi}$, the Cantor set spectrum can be approximated to consist of finitely many larger (adiabatic) gaps with $(\Delta E) ^2 \gg \hbar 2 V\dot{\varphi}$ that are traversed adiabatically during the pumping, and infinitely many smaller (diabatic) gaps with $(\Delta E) ^2 \ll \hbar 2 V\dot{\varphi}$ that are irrelevant for the pumping dynamics. Hence only the adiabatic gaps need be considered. 
For decreasing pump rates, more and more gaps become adiabatic and the effective bands  split into new bands indexed by typically larger Chern numbers. For instance, once the 2nd-order gaps become adiabatic, the central $C=-2$ 1st-order band splits into three bands with $C = -3\,,+4\,,-3$, which together add up to $C=-2$ (see \cref{fig:fig1} c). 

The above division into adiabatic and diabatic gaps is particularly powerful in the regime $ V/J   \gg  1$, where the separation in energy gaps $\Delta E^{(n)}/\Delta E^{(n+1)}$ is largest. Moreover, in this regime, Anderson localization permits an effective model (with exponential precision in $V/J$) in which eigenstates are completely localized on a single site. 
In this regime, the configuration space representation offers a convenient geometric view of adiabatic pumping, see \cref{fig:fig_pumping}: an increase in $\varphi$ acts as a clockwise rotation in configuration space (\cref{fig:fig_pumping} c and d) and by continuously varying $\varphi$, every single-site localized state eventually reaches a resonance, $\theta^{(n)}$, where it resonates with another state localized $n$ sites away. The dynamics of the two resonant sites is well approximated by an effective 2-by-2 Hamiltonian
\begin{equation}
    H_{2\times2}^{(n)}(t) = \begin{pmatrix}
       \delta^{(n)}(t)/2 & J_{eff}^{(n)} \\
        J_{eff}^{(n)}& -\delta^{(n)}(t)/2 
    \end{pmatrix},
\end{equation}
whose detuning 
\begin{align}\label{eq:delta_n}
    \delta^{(n)}(t) = 2V&\bigg[\cos{\left(\theta^{(n)}+\varphi(t) \right)}\nonumber\\
    &-\cos{\left(\theta^{(n)}+n2\pi\beta+\varphi(t) \right)}\bigg]
\end{align}
varies continuously, with one site going up in energy while the other goes down with $\varphi$, see the inset \cref{fig:fig1}a. If the gap is adiabatic ($(\Delta E) ^2 \gg \hbar 2 V\dot{\varphi}$), i.e., the pump rate $\dot{\varphi}$ is slow compared to the corresponding tunneling process $J_{eff}^{(n)} \propto J^{n}/V^{n-1}$, the state adiabatically follows an eigenstate of $H_{2\times2}^{(n)}$ and undergoes a Landau-Zener transition to the resonant $n$-th order neighbor -- giving rise to the quantized transport. In configuration space, this Landau-Zener tunnelling translates into a discrete jump of $ n 2\pi \beta$.

The quantized current for a given filled band (in the case $\beta >1/2$) can be inferred as follows:
In the upper first-order band, indicated by the shaded sector in \cref{fig:fig_pumping} d, states are rotated along the blue arrow and will undergo a Landau-Zener transitions every time the phase winds by  $\Delta\varphi = 2\pi(1-\beta)$. 
Since in this band the transition is to the neighboring site $(n=1)$,  every state is transported by on average $1/(1-\beta)$ sites per pump cycle.
Combined with the fraction $(1-\beta)$ of states in this band,  the total pumped charge per pump cycle in this band is $C_p = (1-\beta)\frac{1}{1-\beta} = +1$, from which we extract $C = +1$. The cases where $\beta<1/2$ can be treated by sending $\beta \rightarrow 1-\beta$. 
The Chern number of any other filled band $B^{(m,n)}$ can be extracted analogously from this local picture by assuming that the band contains only diabatically small gaps and is enclosed by two adiabatically large spectral gaps.

Finally, this configuration-space picture directly confirms that the Chern number (gap label) of an $n$-th order gap $\Delta E^{(n)}$, i.e.\ the quantised current arising from filling all states up to this gap, must be equal to $\pm n$, since the fraction of states below a given gap is exactly equal to the phase $\Delta \varphi$ between consecutive Landau-Zener transitions, which  will transport the state by $n$ sites. For determining the sign we refer the reader to \cref{app:gaplabels}.

The configuration-space picture matches an alternative, global computation of the quantized current using the smooth projector
\begin{align}
    \hat{P}^{(m,n)}(\varphi) = \oint_{\gamma_{m,n}(\varphi)} dz \left(z-\hat{H}^{J,V}_{AAH}(\varphi)\right)^{-1}\,,
\end{align}
 see Ref.~\cite{Kato1950} for further details. In the absence of disorder (see below), the adiabatic transport resulting from both the projector and the local Landau-Zener picture are equivalent by Theorem 1.2 in Ref.~\cite{hagedorn1991proof}.
 However, the proof of quantized currents in the projector picture depends on a global condition, namely the existence of sufficiently large spectral gaps in the spectrum of $\hat{H}^{J,V}_{AAH}(\varphi)$. By contrast, the Landau-Zener transitions in the configuration-space picture depend only on local resonance conditions. This key difference allows the configuration-space picture to be extended to strong disorder.

\begin{figure*}
    \centering
    \includegraphics[width = \linewidth]{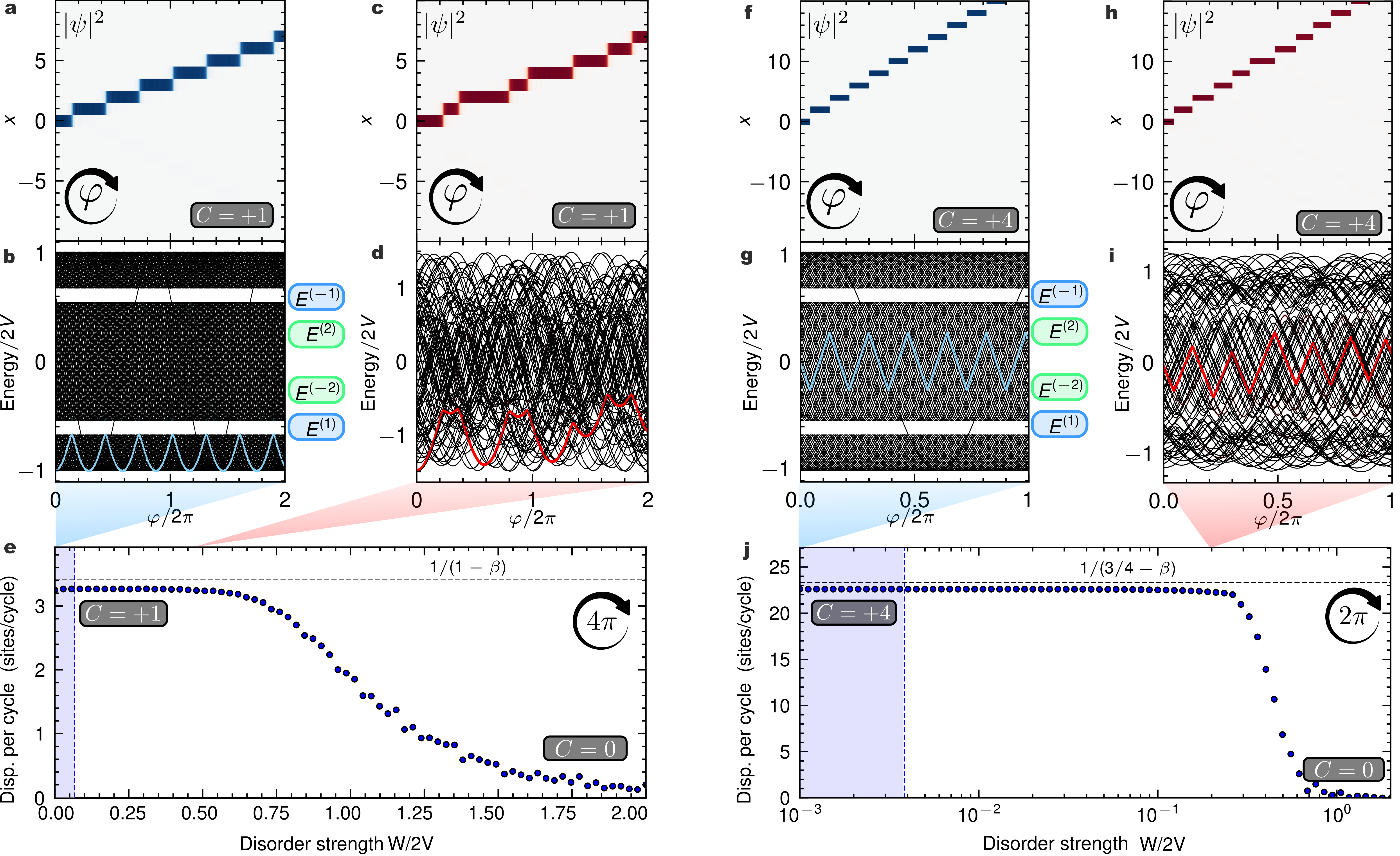}
    \caption{\textbf{Stability and breakdown of quantized currents under onsite disorder:}
    Left side shows results for the lower $C=+1$ first-order band, while the right side depicts the $C=+4 $ second-order band.
    (\textbf{a,f}) Thouless pumping of a particle in the clean case $W=0$, resulting in a quantized average displacement per cycle. \textbf{ (b,g)}  The particle stays confined within its energy band surrounded by open gaps throughout the Thouless pump. Blue curve shows the projection of the wavefunction onto the instantaneous eigenstates plotted in black. (\textbf{c,h})  In the disordered case, the averaged displacement remains quantized, despite the energy gaps being closed (\textbf{d,i}) by the disorder. (\textbf{e,j}) Average displacement per pump cycle after $2$ (\textbf{e}) or $1$ (\textbf{j}) pump periods, showing quantisation also past the closing of the clean energy gaps (blue regions).   The  critical disorder strength for $C=+4$ is more than an order of magnitude larger than the clean gaps. Horizontal dotted lines indicate the current expected in the clean limit for an infinite pump period. Each point is obtained after averaging over $1000$ disorder realisations. \textit{Parameters}: $\beta = \sqrt{2}/2$, $J = 1$, $V = 7.5$, $\dot{\varphi} = 5 \cdot 10^{-2}$ for $C=+1$, and $\dot{\varphi} = 10^{-4}$ for $C=+4$.}
    \label{fig:fig_breakdown}
\end{figure*}

\section{Robustness to spatial disorder}\label{section:disorder}
We will now add local disorder
\begin{align}
    \hat{W} = \sum_{i} w_{i} \hat{a}^{\dagger}_{i} \hat{a}_{i},
\end{align}
where the $w_{n}\in [-W,W]$ are independent, identically distributed random variables. The full disordered Hamiltonian is then 
\begin{align}\label{eq:AAH_dirty}
	\hat{H}^{J,V,W} (\varphi)= \hat{H}_{AAH}^{J,V}(\varphi)+\hat{W}.
\end{align}
Its eigenfunctions are all Anderson localized both by the disorder $\hat{W}$ \cite{PhysRev.109.1492} and, for $V>J$, by the famous Aubry-Andr\'e duality mentioned above~\cite{jitomirskaya1998anderson}.

It was shown in~\cite{avilanewresult} that the resulting spectrum $\sigma\left(\hat{H}^{J,V,W} (\varphi)\right)$ is the span of the pairwise sum of elements in the unperturbed spectrum $\sigma\left(\hat{H}^{J,V}_{AAH}(\varphi)\right)$ with elements from the disorder distribution $[-W,W]$,
\begin{align}\label{eq:disorderedspectrum}
    \sigma(\hat{H}^{J,V,W} (\varphi)) =  \sigma(\hat{W}) \star \sigma\left(\hat{H}^{J,V}_{AAH}(\varphi)\right).
\end{align}
Intuitively, this means that the disordered energy spectrum can be obtained by convolving the clean spectrum with the spectrum of the disorder term. Therefore, it follows that the disorder closes all unperturbed spectral gaps that are smaller than the bandwidth of the disorder strength $\Delta E^{(n)} < 2W$.

We numerically compute the pumped charge as the average displacement, that is the average number of sites traveled, per pump cycle from a linear fit to the center of mass motion for a single particle initially localized at a configuration-space position which corresponds to a band with Chern number $C = +1$ (\cref{fig:fig_breakdown} a) and $+4$ (\cref{fig:fig_breakdown} f) in the clean case $W=0$. For small disorders below the closing of the relevant gaps (blue shaded area), the extracted values agree as expected with the above quantized predictions in the presence of global gaps, see \cref{fig:fig_breakdown}~e,~j. However, the quantized currents survive far beyond the closing of the clean energy gaps (dashed blue lines), and break down only at larger critical disorder strengths.   In the $C=+4$ band shown in \cref{fig:fig_breakdown} j, we find that the critical disorder strength for this particular band  exceeds the size of the clean $\Delta E^{(\pm 2)}$ gaps by more than an order of magnitude. 

\begin{figure}
    \centering
    \includegraphics[width = \linewidth]{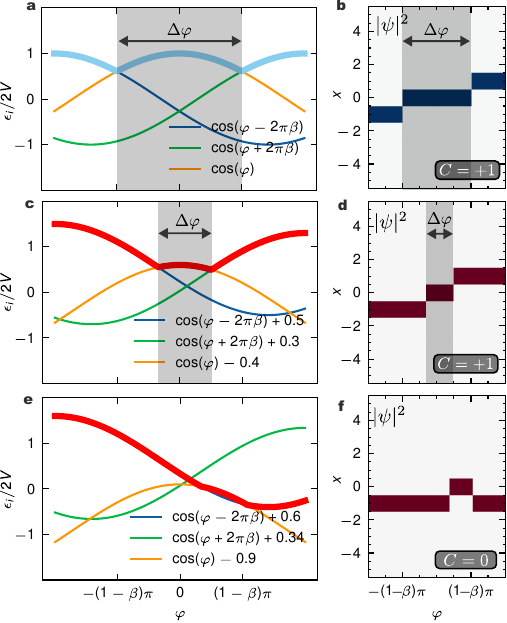}
    \caption{\textbf{Local picture for the stability and breakdown of quantized currents:}
    Coloured lines in the left column panels denote onsite energies of three consecutive sites for $V\gg J$. Thick coloured line illustrates the adiabatic trajectory of the state under pumping. The right column shows the corresponding real-space density. (\textbf{a,b}) In the clean case, the phase $\Delta \varphi$ separating two consecutive Landau-Zener transitions is the same for each transition and the particle undergoes successive adiabatic transitions to neighboring sites at a constant rate (blue curve). (\textbf{c,d}) The addition of not-too-large onsite disorder induces fluctuations in the phase $\Delta \varphi$ between transitions but preserves the order of resonances and leaves the pumped current quantized. (\textbf{e,f}) When the disorder strength exceeds a critical value, the order of successive Landau-Zener processes can change, causing the particle to be back-reflected. The argument is shown here for the top $C=+1$ first-order band. \textit{Parameters:} $\beta = \sqrt{2}/2$.}
   
    \label{fig:fig_disorder_resonances}
\end{figure}

Once more, configuration space provides a simple picture for the stability of the quantized current. In the clean case, the phase winding $\Delta \varphi$ between two consecutive Landau-Zener processes for a given localized state is constant (\cref{fig:fig_disorder_resonances} a,b) and corresponds to the angle spanned by the corresponding band on the unit circle, see \cref{fig:fig_pumping}d. 
In the presence of disorder, this phase fluctuates randomly, giving rise to a distribution of spatially-dependent and randomly distributed $\Delta \varphi_i = \Delta \varphi + \delta \varphi_i$. This is reflected in the fluctuating size of the plateaus in \cref{fig:fig_breakdown} c.  As long as the disorder is below a critical value, these fluctuations $\delta \varphi_i$ average to zero and leave the quantized current unaffected (\cref{fig:fig_disorder_resonances} c,d). Once the disorder becomes strong enough to shrink one ore more of the $\Delta \varphi_i$ down to $0$ (\cref{fig:fig_disorder_resonances} e,f), the order of resonances may change, which results in the breakdown of the quantized current as the particle starts being back-reflected. 

\begin{figure}
    \centering
    \includegraphics[width=\linewidth]{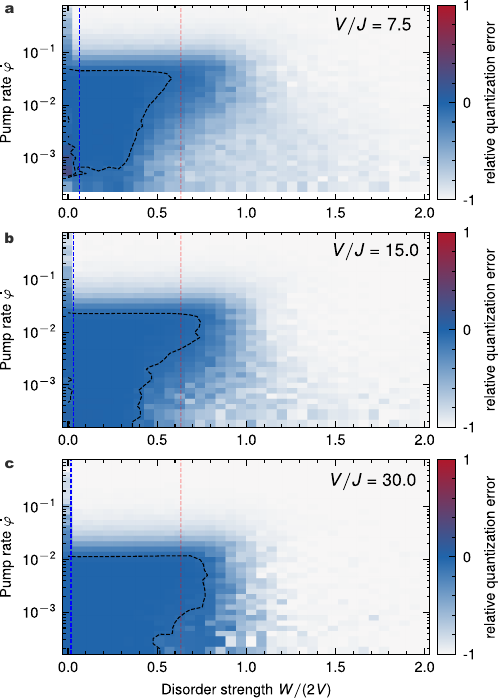}
    \caption{\textbf{Effect of pumping rate on the quantization of the current in the $C=+1$ band:} Color denotes the average relative error in the quantization of the displacement per cycle (extracted from the slope of the right-moving part of the density $|\Psi(x>0)|^2$ of an initial state consisting of a product of $20$ local Fock states on neighboring sites, cf.\ \cref{fig:pumping_cloud}). Black contour lines indicate the regions with less than $5 \%$ error. Blue dotted line indicates the closing of the first-order energy gaps due to disorder. Red dotted line indicates the configuration-space estimate for the critical disorder strength (\cref{eq:critical_c_1_main}).
    For very low pumping rates, current stops being quantized once unwanted longer-range tunneling processes become relevant. These additionally  increase the sensitivity of the pump to the disorder. The increasing  separation of timescales between the first- and second-order gaps for larger $V/J$ allows for a better protection of the quantization for large $V/J$.   \textit{Parameters:} (a) $V/J = 7.5$. (b) $V/J = 15.0$. (c) $V/J = 30.0$. Pumped angle is $15\pi$. Each point is obtained after averaging over $20$ disorder realizations.}
    \label{fig:rate_critical_disorder}
\end{figure}

The above analysis rests on the assumption that the pumping rate is high enough to prevent the occurrence of Landau-Zener tunneling over longer ranges than desired. To illustrate this necessary condition, we investigate the impact of the pumping rate on the critical disorder strength  for destroying the quantized currents in the $C=+1$ bands. \cref{fig:rate_critical_disorder} shows the relative error in the quantization of the displacement per cycle of the right-moving part of an initial state consisting of a product of $20$ local Fock states on neighboring sites, see \cref{fig:pumping_cloud} for reference. The red dotted line shows the configuration-space estimate
\begin{equation}\label{eq:critical_c_1_main}
	W_{crit}^{C = +1} = 2V\sin^2{\left(\pi \beta\right)}.
\end{equation}
of the disorder strength necessary for the condition $\Delta \varphi_i=0$, see details in \cref{appendix:disorder}. 

The quantized current emerges for pumping rates low enough to be adiabatic with respect to the largest spectral gaps of order $J$, i.e., for $\dot{\varphi}\approx\mathcal{O}(2J^2/V)$. The quantization breaks down again when the pumping becomes adiabatic with respect to the next spectral gap of order $J_{\textit{eff}}^{(2)}$, i.e., $\dot{\varphi}\approx\mathcal{O}(2J^{4}/V^3)$, and hence enables undesired longer-range tunneling. Consistent with \citep{vuina2024absence}, the protection of the quantization to disorder closing spectral gaps is not expected to hold in the full adiabatic limit, where Landau-Zener transitions can take place over arbitrary distances. Interestingly, we observe that for slow pumping the addition of weak disorder induces a stabilization of the quantization, owing to the suppression of pumping via longer-range resonances. The critical disorder strength observed in \cref{fig:rate_critical_disorder}, where the initial state is a product state of many particles, is in good agreement with the single-particle results shown in \cref{fig:fig_breakdown} e.

Thus, provided that the pumping rate is set correctly to suppress undesired longer-range tunneling processes, the robustness of the quantized current is not limited by the size of the clean energy gaps, but instead depends on the ordering of consecutive local resonances. In the clean case, these resonances are all aligned in energy, giving rise to global gaps. Increasing onsite disorder shifts each resonance independently  and thereby  closes the global energy gaps, but does not necessarily modify their ordering. As long as the order between consecutive local resonances remains unaffected, quantized currents can subsist, even in the complete absence of global energy gaps in the system.

The above robustness is distinct from the situation in conventional disordered topological systems studied in the context of the integer and fractional quantum hall effects~\cite{PhysRevB.23.5632}, where the robustness beyond the closing of the spectral gap stems from the persistence of a mobility gap between the (gapped) topological extended and (gapless) localized states of the system~\citep{articleshapiro} and is often described using real-space topological invariants called Chern markers~\cite{PhysRevB.84.241106}. 
However, this picture does not directly apply to the 1D AAH Hamiltonian as it only hosts insulating states for $V>J$. 
At first glance, the notion of extended states may be restored by treating $\varphi$ as a virtual dimension in $\hat{H}^{J,V,W} (\varphi)$ and applying the famous mapping of the 1D AAH Hamiltonian to the 2D Hofstadter Hamiltonian~\citep{aubry1980annals}, see \cref{sec:StatePrep}. However, the onsite disorder $\hat{W}$  considered here only generates a striped disorder pattern in 2D that cannot localize states along the virtual dimension~\cite{Bordia2016}. 
Thus, all states are extended and the traditional mobility gap closes.
 
Interestingly, the Landau-Zener formalism above implicitly depends on the finite pumping period $T$ to limit long-range resonances in the presence of strong disorder~\citep{vuina2024absence}, see \cref{fig:rate_critical_disorder}. This period also appears in the 1D to 2D mapping when considering the full Schr\"odinger operator, $\hat{K} \equiv i\partial_{t} - \hat{H}^{J,V,W}(\varphi(t))$. 
Since $\hat{H}^{J,V,W} (\varphi(t))$ is periodic in time with period $T$, only discrete frequencies $\omega=\dot{\varphi} \tilde{j}$ with $\tilde{j}\in \mathbb{Z}$ appear when
Fourier transforming $\hat{K}$ along the time coordinate, forming a discrete virtual  dimension~\citep{Ozawa2019, prodanVirtual2015}.  The Fourier transform furthermore exchanges the time-derivative for a linear potential proportional to the pumping rate $\dot{\varphi}$, $\tilde{K} \equiv \sum_{\tilde{j}\in \mathbb{Z}} \dot{\varphi}\tilde{j} - \hat{H}_{QHE}-\hat{W} \otimes \mathbb{I}_{\tilde{j}}$ \citep{https://doi.org/10.1002/cpa.3160070404}, see \cref{eq:2dlandau} for definition of $\hat{H}_{QHE}$. 
The eigenfunctions of the Schr\"odinger operator, i.e.\ those states satisfying $\tilde{K}\psi_{\lambda} = \lambda\psi_{\lambda}$, form a diagonal basis for the unitary $\hat{U}(T)$ describing the stroboscopic time evolution of states in the 1D Thouless pump. Thus, a possible reconciliation is that the finite pumping rate introduces a non-vanishing electric field (the linear potential) along the Fourier-transformed virtual $\tilde{j}$ direction and induces Wannier-Stark localization along this direction~\citep{Wannier1960432}. The mix of Wannier-Stark and Anderson localization localizes topologically trivial states and generates a hybrid mobility gap. This suggests the existence of finer definitions for mobility gaps in the Thouless pumping of quasiperiodic systems. Determining whether or not these two pictures are indeed equivalent constitutes a promising direction for future work.

\section{Experimental Signatures and Quantum State Preparation}
\label{sec:StatePrep}

The robustness of the quantized currents can not only be directly observed experimentally, but can also be leveraged to turn the Thouless pump into a robust protocol for preparing quantum states with desired Chern numbers. 
We propose here an instance of the protocol that is directly implementable within existing cold atom experiments. 

The one-dimensional AAH model can be realised by superimposing one strong (primary) optical lattice with another weaker incommensurate lattice~\cite{Roati2008, schreiber,nakajimaCompetitionInterplayTopology2021b, marraTopologicallyQuantizedCurrent2020}, where deep lattices in the transverse directions confine the atoms to 1D chains. The depth of the primary lattice sets $J$ and the depth of the secondary lattice depth provides control over $V/J$. The same model can also be realized by independently controlling the depths of the individual lattices in an optical quasicrystal~\cite{gottlobPRB, sbrosciaQC, ViehbanQC}. For large systems, the initial phase $\varphi$ of the quasiperiodic modulation is irrelevant. 
The initial state can be prepared for instance as a unit-filling band insulator of spinless fermions or hard-core bosons in a projected optical box potential \cite{Navon2021}. Such a state naturally samples the energy spectrum uniformly.

In contrast to periodic systems, the AAH system can be realised in the regime $V\gg J$, where all eigenstates are strongly localized. In this regime, the only motion stems from the pumping, and the atomic cloud will not undergo any free expansion, such that no additional trap is required.

\begin{figure}
    \centering
    \includegraphics[width = \linewidth]{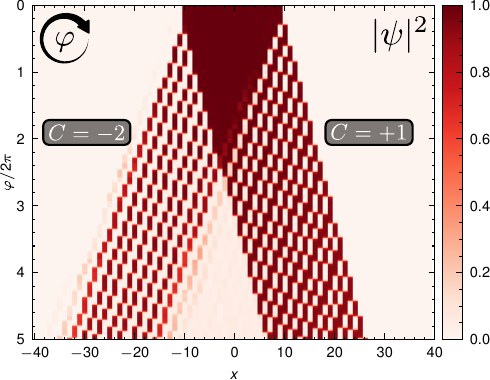}
    \caption{\textbf{Preparing states with desired Chern numbers:} The initial state is prepared as a product of local Fock states with unit filling and the pumping rate $\dot{\varphi}$ is chosen to be adiabatic with respect to the first-order gaps only. The atom cloud splits into a left- and a right-moving part that move at different speeds, separating the middle band  with Chern number $C=-2$ from the top and bottom bands with $C=+1$. The desired Chern number can be selected by e.g.\ eliminating the remaining other branches with a resonant laser pulse.  \textit{Parameters:} $\beta = \sqrt{2}/2$,
    $J=1$, $V = 7.5$, $W/(2V) = 0.3$. $\dot{\varphi} = 5 \cdot 10^{-2}$.}  \label{fig:pumping_cloud}
\end{figure}

States in different bands become spatially separated as $\varphi$ is pumped. For example, if the pumping rate is set to be adiabatic with respect to the first-order gaps only, the Thouless pump splits the cloud in two parts (\cref{fig:pumping_cloud}). The right moving part corresponds to atoms populating the lower and upper bands with $C = +1$, while the left-moving part contains the atoms in the middle band with $C = -2$. Note that the center of mass position of the entire cloud remains constant at all times, since the total Chern number of the entire spectrum vanishes ($C=0$). To prepare a state with the desired Chern number, it then suffices to select the desired branch by clearing  out atoms in the (spatially separated) undesired branches by, for example, exposing them to resonant laser light.

\begin{figure} 
    \centering
    \includegraphics[width = \linewidth]{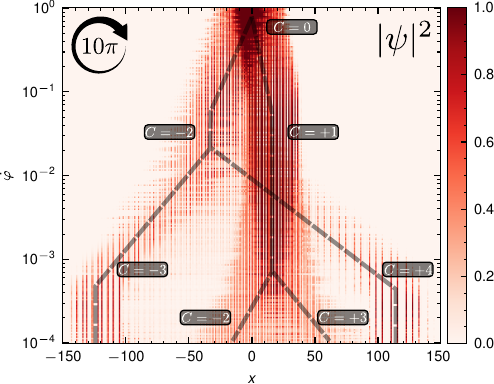}
    \caption{\textbf{Effect of pumping rate:} Final  density after 5 pump cycles for various pump rates $\dot{\varphi}$. The initial state is a Fock state with one atom per site within a selected region. By decreasing the pump rate $\dot{\varphi}$, the Thouless pump becomes sensitive to smaller gaps, characterised by increasingly higher Chern numbers. This is reflected in the successive branchings of the final density as the pumping rate is decreased, highlighted by the dashed guidelines. \textit{Parameters:}  $\beta = \sqrt{2}/2$, $J=1$, $V=5 $,\ $W/(2V) = 0.01$.} \label{fig:pumping_rate}
\end{figure}

The Thouless pump can be made sensitive to increasingly smaller gaps, and hence higher Chern numbers, by reducing the pumping rate $\dot{\varphi}$, leading to  successive branchings of the final density, see \cref{fig:pumping_rate}. The robustness of the Thouless pump to onsite disorder ensures that this state preparation protocol is robust even when the energy gaps surrounding the targeted energy bands are small. This protocol can be directly extended to the interacting regime by using fermionic spin mixtures or real bosons and tuning the onsite interaction strength via a Feshbach resonance \cite{ChinFeshbach}.

The above procedure generates a 1D many-body state with an associated virtual Chern number. To generate a state with a well-defined ``real" 2D Chern number, the protocol can in principle be extended to a 2D lattice consisting of an array of $n_y$ AAH chains with increasing phase $\varphi(j) = 2\pi j/n_y$ along a second spatial dimension $y$. This results in a  2D Hamiltonian given by 
\begin{align}\label{eq:2dstacked}
    \hat{H}_{2D} = \sum_{(i,j)} &2V\cos\left( 2\pi\beta i+\frac{2\pi}{n_y}j\right) \hat{a}_{i,j}^{\dagger}\hat{a}_{i,j} \nonumber \\
    &-J\left(\hat{a}_{i+1,j}^{\dagger}\hat{a}_{i,j} +\hat{a}_{i,j}^{\dagger}\hat{a}_{i+1,j}\right)\,,
\end{align} 
whose spectrum is the same as that of the 1D AAH Hamiltonian \cite{PhysRevB.14.2239,aubry1980annals}. Under Fourier transforming $j\rightarrow\tilde{j}$, \cref{eq:2dstacked} is equivalent to the 
2D Harper Hamiltonian 
\begin{align}\label{eq:2dlandau}
    \hat{H}_{QHE} = \sum_{(i,\tilde{j})}&V\left( e^{i2\pi\beta i}\hat{a}_{i,\tilde{j}+1}^{\dagger}\hat{a}_{i,\tilde{j}} +h.c.\right)\nonumber \\ &- J\left(\hat{a}_{i+1,\tilde{j}}^{\dagger}\hat{a}_{i,\tilde{j}} +h.c.\right)
     ,
\end{align}
which describes a square lattice with magnetic flux $2\pi\beta$ per plaquette.
Therefore, by stacking multiple copies of the one-dimensional state preparation protocol, we prepare a state with a desired Chern number of the 2D Harper Hamiltonian, albeit in a mixed real-momentum space configuration. 
The chains (momentum classes) can be coupled by introducing small tunneling elements between neighboring chains without altering the Chern number of the state, provided that the spectral gaps remain open \cite{clas1b}. 

Interestingly, onsite interactions  effectively appear as local momentum-space interactions along the $\tilde{j}$ coordinate \citep{PhysRevLett.120.040407}. This mixed real-momentum space system presents novel opportunities to probe new quantum Hall physics inaccessible to traditional condensed-matter systems by for example engineering  scalar potentials in momentum space to reproduce dispersion relations corresponding to long-range real-space hopping models that have been used to generate exact fractional quantum hall (FQH) phases on a lattice via topological flat bands \citep{PhysRevB.88.205101}. Similarly, non-local real-space interactions have been used to stabilize non-abelian fractional Chern insulator (FCI) phases in numerics \citep{PhysRevLett.105.215303}. Moreover, numerical studies of fractional Chern insulators have long noticed enhanced stability of the FCI phase in higher Chern number bands \citep{PhysRevB.86.205125} along with new phases of matter with no FQH analogs \citep{PhysRevLett.109.186805}. As such, the above state preparation protocol could become a new experimental platform for exploring numerous theoretical proposals involving large Chern numbers, flat bands, and non-local interactions.

\section{Discussion and Conclusion}
We present a novel example of topological protection in quasiperiodic systems that does not depend on the presence of spectral gaps \cite{Kato1950}, symmetry protections \cite{PhysRevX.7.041048}, or even traditional mobility gaps~\cite{PhysRevB.23.5632}. By tracking the fate of quantized currents in the disordered Aubry-André chain (1D almost Mathieu operator), we find that the combination of quasiperiodic Anderson localization and the gap labeling theorem results in a robustness to bounded local disorder far beyond the closing of the relevant spectral gaps. We furthermore develop a local picture to explain quasiperiodic Thouless pumping both in the clean and disordered case to explain the observed robustness. 

As mentioned in the main text, determining whether or not the local picture developed here and the global breakdown of mixed linear/disorder potential mobility gaps are indeed equivalent constitutes a promising direction for future work. In particular, for weak disorder, where (large) spectral gaps of the 1D AAH Hamiltonian remain open, the infinite-range quasiperiodic resonances do not obstruct quantized pumping in low-order bands as $\dot{\varphi}\rightarrow 0$, despite the continued bifurcations into smaller bands. For strong disorder, on the other hand,  finite  pump frequencies play an important role in protecting the quantized currents by truncating long-range Landau-Zener transitions and possibly establishing a hybrid mobility gap, consistent with the picture in Ref.~\citep{vuina2024absence}. This change in the role of long-range resonances with increasing disorder suggests a potentially interesting region of parameter space for future study where the character of the 1D eigenstates is preserved, but the topological invariants of the 2D Hamiltonian are not protected by a mobility gap. 
Furthermore, coupling many of these 1D chains into a 2D system presents novel opportunities to probe new quantum Hall physics inaccessible to traditional condensed-matter systems.

The outlined mathematical relations to ergodicity and the Dry Ten Martini problem \cite{avila2009ten,han2018dry,drytenmartininoncritical} present another direction for future work. It would for instance be worth exploring the case of several incommensurate frequencies, which are likely to result in enhanced stabilization mechanisms that in turn should lead to novel measurable signatures. These pursuits can also be followed in the context of the Dry Ten Martini problem without dualities or with hidden dualities~\cite{han2018dry, drytenmartininoncritical,  goncalves2021hidden}, where localization phase transitions may be richer. 

In two-dimensional quasiperiodic systems, i.e.\ systems that are quasiperiodic along both dimensions, the presence of second Chern numbers and a richer set of topological invariants could lead to more examples of gap-independent topological protection.
For example, in ultracold atomic systems, the scheme could be extended to the two-dimensional Aubry-André model or optical quasicrystals~\citep{gottlobPRB, sbrosciaQC, ViehbanQC}, where pumping could be realised along multiple dimensions.
The recent advent of moir\'e van-der-Waals structures, in which pumping can be achieved upon applying periodic external gauge fields, provides an exciting solid-state platform in which these new mechanisms might both occur and help explain experimental observations.

In summary, our results uncover a novel protection mechanism for Thouless pumping in quasiperiodic systems, highlighting an unexplored richness of topology in quasiperiodic systems and motivating a re-examination of topological protection in the absence of translation invariance. They furthermore enable novel experimental applications, notably in the preparation of quantum states with high Chern numbers.

\section{Acknowledgements}
We cordially thank Joel E.\ Moore for insightful feedback and advice regarding the draft. We also thank Arjun Ashoka, Alexander Avdoshkin, Vir B. Bulchandani, Fabian Heidrich-Meisner, Matthew Henricks, Johannes Mitscherling, Suman Mondal, Lee Reeve, Ruben Verresen, Qijun Wu, and Maciej Zworski for many helpful discussions.
R.-J.S.\ acknowledges funding from a New Investigator Award, EPSRC grant EP/W00187X/1, a EPSRC ERC underwrite grant  EP/X025829/1, and a Royal Society exchange grant IES/R1/221060 as well as Trinity College, Cambridge.
Work by D.S.B.\ was
performed with support through a MURI project supported by the Air Force Office of Scientific
Research (AFSOR) under grant number FA9550-22-1-027.
U.S.\ and E.G.\ acknowledges support  by the European Commission ERC Starting Grant QUASICRYSTAL, the
EPSRC Grant (No. EP/R044627/1), and EPSRC Programme Grant DesOEQ (No. EP/P009565/1) and the Deutsche Forschungsgemeinschaft (DFG, German Research Foundation) via Research Unit FOR 2414 under project number 277974659. E.G. acknowledges support from the Cambridge Trust.
\section{Authors contributions}
E.~G.\ with input from U.~S.\  set up and performed the numerical and analytical configuration-space analysis for the Thouless pump. Relations to mathematical spectral theory and the discovery of the underlying stability mechanism under disorder progressed with insights from D.~S.~B.\ and R.-J.~S.. D.~S.~B. had a steering role in turning to the stability under disorder. All aspects were extensively discussed amongst all authors. All authors contributed to the writing of the manuscript. 

\newpage
\bibliography{biblio}

\begin{thebibliography}{61}%
\makeatletter
\providecommand \@ifxundefined [1]{%
 \@ifx{#1\undefined}
}%
\providecommand \@ifnum [1]{%
 \ifnum #1\expandafter \@firstoftwo
 \else \expandafter \@secondoftwo
 \fi
}%
\providecommand \@ifx [1]{%
 \ifx #1\expandafter \@firstoftwo
 \else \expandafter \@secondoftwo
 \fi
}%
\providecommand \natexlab [1]{#1}%
\providecommand \enquote  [1]{``#1''}%
\providecommand \bibnamefont  [1]{#1}%
\providecommand \bibfnamefont [1]{#1}%
\providecommand \citenamefont [1]{#1}%
\providecommand \href@noop [0]{\@secondoftwo}%
\providecommand \href [0]{\begingroup \@sanitize@url \@href}%
\providecommand \@href[1]{\@@startlink{#1}\@@href}%
\providecommand \@@href[1]{\endgroup#1\@@endlink}%
\providecommand \@sanitize@url [0]{\catcode `\\12\catcode `\$12\catcode `\&12\catcode `\#12\catcode `\^12\catcode `\_12\catcode `\%12\relax}%
\providecommand \@@startlink[1]{}%
\providecommand \@@endlink[0]{}%
\providecommand \url  [0]{\begingroup\@sanitize@url \@url }%
\providecommand \@url [1]{\endgroup\@href {#1}{\urlprefix }}%
\providecommand \urlprefix  [0]{URL }%
\providecommand \Eprint [0]{\href }%
\providecommand \doibase [0]{https://doi.org/}%
\providecommand \selectlanguage [0]{\@gobble}%
\providecommand \bibinfo  [0]{\@secondoftwo}%
\providecommand \bibfield  [0]{\@secondoftwo}%
\providecommand \translation [1]{[#1]}%
\providecommand \BibitemOpen [0]{}%
\providecommand \bibitemStop [0]{}%
\providecommand \bibitemNoStop [0]{.\EOS\space}%
\providecommand \EOS [0]{\spacefactor3000\relax}%
\providecommand \BibitemShut  [1]{\csname bibitem#1\endcsname}%
\let\auto@bib@innerbib\@empty
%</preamble>
\bibitem [{\citenamefont {Thouless}\ \emph {et~al.}(1982)\citenamefont {Thouless}, \citenamefont {Kohmoto}, \citenamefont {Nightingale},\ and\ \citenamefont {den Nijs}}]{ThoulessQHE}%
  \BibitemOpen
  \bibfield  {author} {\bibinfo {author} {\bibfnamefont {D.~J.}\ \bibnamefont {Thouless}}, \bibinfo {author} {\bibfnamefont {M.}~\bibnamefont {Kohmoto}}, \bibinfo {author} {\bibfnamefont {M.~P.}\ \bibnamefont {Nightingale}},\ and\ \bibinfo {author} {\bibfnamefont {M.}~\bibnamefont {den Nijs}},\ }\bibfield  {title} {\bibinfo {title} {Quantized hall conductance in a two-dimensional periodic potential},\ }\href {https://doi.org/10.1103/PhysRevLett.49.405} {\bibfield  {journal} {\bibinfo  {journal} {Physical Review Letters}\ }\textbf {\bibinfo {volume} {49}},\ \bibinfo {pages} {405} (\bibinfo {year} {1982})}\BibitemShut {NoStop}%
\bibitem [{\citenamefont {Halperin}(1982)}]{halperin1982quantized}%
  \BibitemOpen
  \bibfield  {author} {\bibinfo {author} {\bibfnamefont {B.~I.}\ \bibnamefont {Halperin}},\ }\bibfield  {title} {\bibinfo {title} {Quantized hall conductance, current-carrying edge states, and the existence of extended states in a two-dimensional disordered potential},\ }\href@noop {} {\bibfield  {journal} {\bibinfo  {journal} {Physical Review B}\ }\textbf {\bibinfo {volume} {25}},\ \bibinfo {pages} {2185} (\bibinfo {year} {1982})}\BibitemShut {NoStop}%
\bibitem [{\citenamefont {Thouless}(1983)}]{thoulessQuantizationParticleTransport1983}%
  \BibitemOpen
  \bibfield  {author} {\bibinfo {author} {\bibfnamefont {D.~J.}\ \bibnamefont {Thouless}},\ }\bibfield  {title} {\bibinfo {title} {Quantization of particle transport},\ }\href {https://doi.org/10.1103/PhysRevB.27.6083} {\bibfield  {journal} {\bibinfo  {journal} {Physical Review B}\ }\textbf {\bibinfo {volume} {27}},\ \bibinfo {pages} {6083} (\bibinfo {year} {1983})}\BibitemShut {NoStop}%
\bibitem [{\citenamefont {Citro}\ and\ \citenamefont {Aidelsburger}(2023)}]{citroThoulessPumpingTopology2023b}%
  \BibitemOpen
  \bibfield  {author} {\bibinfo {author} {\bibfnamefont {R.}~\bibnamefont {Citro}}\ and\ \bibinfo {author} {\bibfnamefont {M.}~\bibnamefont {Aidelsburger}},\ }\bibfield  {title} {\bibinfo {title} {Thouless pumping and topology},\ }\href {https://doi.org/10.1038/s42254-022-00545-0} {\bibfield  {journal} {\bibinfo  {journal} {Nature Reviews Physics}\ }\textbf {\bibinfo {volume} {5}},\ \bibinfo {pages} {87} (\bibinfo {year} {2023})}\BibitemShut {NoStop}%
\bibitem [{\citenamefont {Kraus}\ \emph {et~al.}(2012)\citenamefont {Kraus}, \citenamefont {Lahini}, \citenamefont {Ringel}, \citenamefont {Verbin},\ and\ \citenamefont {Zilberberg}}]{kraus2012topological}%
  \BibitemOpen
  \bibfield  {author} {\bibinfo {author} {\bibfnamefont {Y.~E.}\ \bibnamefont {Kraus}}, \bibinfo {author} {\bibfnamefont {Y.}~\bibnamefont {Lahini}}, \bibinfo {author} {\bibfnamefont {Z.}~\bibnamefont {Ringel}}, \bibinfo {author} {\bibfnamefont {M.}~\bibnamefont {Verbin}},\ and\ \bibinfo {author} {\bibfnamefont {O.}~\bibnamefont {Zilberberg}},\ }\bibfield  {title} {\bibinfo {title} {Topological states and adiabatic pumping in quasicrystals},\ }\href {https://doi.org/10.1103/PhysRevLett.109.106402} {\bibfield  {journal} {\bibinfo  {journal} {Physical Review Letters}\ }\textbf {\bibinfo {volume} {109}},\ \bibinfo {pages} {106402} (\bibinfo {year} {2012})}\BibitemShut {NoStop}%
\bibitem [{\citenamefont {Prodan}(2015{\natexlab{a}})}]{prodan2015}%
  \BibitemOpen
  \bibfield  {author} {\bibinfo {author} {\bibfnamefont {E.}~\bibnamefont {Prodan}},\ }\bibfield  {title} {\bibinfo {title} {Virtual topological insulators with real quantized physics},\ }\href {https://doi.org/10.1103/PhysRevB.91.245104} {\bibfield  {journal} {\bibinfo  {journal} {Physical Review B}\ }\textbf {\bibinfo {volume} {91}},\ \bibinfo {pages} {245104} (\bibinfo {year} {2015}{\natexlab{a}})}\BibitemShut {NoStop}%
\bibitem [{\citenamefont {Ozawa}\ \emph {et~al.}(2019)\citenamefont {Ozawa}, \citenamefont {Price}, \citenamefont {Amo}, \citenamefont {Goldman}, \citenamefont {Hafezi}, \citenamefont {Lu}, \citenamefont {Rechtsman}, \citenamefont {Schuster}, \citenamefont {Simon}, \citenamefont {Zilberberg},\ and\ \citenamefont {Carusotto}}]{Ozawa2019}%
  \BibitemOpen
  \bibfield  {author} {\bibinfo {author} {\bibfnamefont {T.}~\bibnamefont {Ozawa}}, \bibinfo {author} {\bibfnamefont {H.~M.}\ \bibnamefont {Price}}, \bibinfo {author} {\bibfnamefont {A.}~\bibnamefont {Amo}}, \bibinfo {author} {\bibfnamefont {N.}~\bibnamefont {Goldman}}, \bibinfo {author} {\bibfnamefont {M.}~\bibnamefont {Hafezi}}, \bibinfo {author} {\bibfnamefont {L.}~\bibnamefont {Lu}}, \bibinfo {author} {\bibfnamefont {M.~C.}\ \bibnamefont {Rechtsman}}, \bibinfo {author} {\bibfnamefont {D.}~\bibnamefont {Schuster}}, \bibinfo {author} {\bibfnamefont {J.}~\bibnamefont {Simon}}, \bibinfo {author} {\bibfnamefont {O.}~\bibnamefont {Zilberberg}},\ and\ \bibinfo {author} {\bibfnamefont {I.}~\bibnamefont {Carusotto}},\ }\bibfield  {title} {\bibinfo {title} {Topological photonics},\ }\href {https://doi.org/10.1103/RevModPhys.91.015006} {\bibfield  {journal} {\bibinfo  {journal} {Rev. Mod. Phys.}\ }\textbf {\bibinfo {volume} {91}},\ \bibinfo {pages} {015006} (\bibinfo {year} {2019})}\BibitemShut {NoStop}%
\bibitem [{\citenamefont {Lohse}\ \emph {et~al.}(2016)\citenamefont {Lohse}, \citenamefont {Schweizer}, \citenamefont {Zilberberg}, \citenamefont {Aidelsburger},\ and\ \citenamefont {Bloch}}]{lohseThoulessQuantumPump2016a}%
  \BibitemOpen
  \bibfield  {author} {\bibinfo {author} {\bibfnamefont {M.}~\bibnamefont {Lohse}}, \bibinfo {author} {\bibfnamefont {C.}~\bibnamefont {Schweizer}}, \bibinfo {author} {\bibfnamefont {O.}~\bibnamefont {Zilberberg}}, \bibinfo {author} {\bibfnamefont {M.}~\bibnamefont {Aidelsburger}},\ and\ \bibinfo {author} {\bibfnamefont {I.}~\bibnamefont {Bloch}},\ }\bibfield  {title} {\bibinfo {title} {A {{Thouless}} quantum pump with ultracold bosonic atoms in an optical superlattice},\ }\href {https://doi.org/10.1038/nphys3584} {\bibfield  {journal} {\bibinfo  {journal} {Nature Physics}\ }\textbf {\bibinfo {volume} {12}},\ \bibinfo {pages} {350} (\bibinfo {year} {2016})}\BibitemShut {NoStop}%
\bibitem [{\citenamefont {Nakajima}\ \emph {et~al.}(2016)\citenamefont {Nakajima}, \citenamefont {Tomita}, \citenamefont {Taie}, \citenamefont {Ichinose}, \citenamefont {Ozawa}, \citenamefont {Wang}, \citenamefont {Troyer},\ and\ \citenamefont {Takahashi}}]{nakajimaTopologicalThoulessPumping2016}%
  \BibitemOpen
  \bibfield  {author} {\bibinfo {author} {\bibfnamefont {S.}~\bibnamefont {Nakajima}}, \bibinfo {author} {\bibfnamefont {T.}~\bibnamefont {Tomita}}, \bibinfo {author} {\bibfnamefont {S.}~\bibnamefont {Taie}}, \bibinfo {author} {\bibfnamefont {T.}~\bibnamefont {Ichinose}}, \bibinfo {author} {\bibfnamefont {H.}~\bibnamefont {Ozawa}}, \bibinfo {author} {\bibfnamefont {L.}~\bibnamefont {Wang}}, \bibinfo {author} {\bibfnamefont {M.}~\bibnamefont {Troyer}},\ and\ \bibinfo {author} {\bibfnamefont {Y.}~\bibnamefont {Takahashi}},\ }\bibfield  {title} {\bibinfo {title} {Topological {{Thouless}} pumping of ultracold fermions},\ }\href {https://doi.org/10.1038/nphys3622} {\bibfield  {journal} {\bibinfo  {journal} {Nature Physics}\ }\textbf {\bibinfo {volume} {12}},\ \bibinfo {pages} {296} (\bibinfo {year} {2016})}\BibitemShut {NoStop}%
\bibitem [{\citenamefont {Walter}\ \emph {et~al.}(2023)\citenamefont {Walter}, \citenamefont {Zhu}, \citenamefont {G\"{a}chter}, \citenamefont {Minguzzi}, \citenamefont {Roschinski}, \citenamefont {Sandholzer}, \citenamefont {Viebahn},\ and\ \citenamefont {Esslinger}}]{Walter2023}%
  \BibitemOpen
  \bibfield  {author} {\bibinfo {author} {\bibfnamefont {A.-S.}\ \bibnamefont {Walter}}, \bibinfo {author} {\bibfnamefont {Z.}~\bibnamefont {Zhu}}, \bibinfo {author} {\bibfnamefont {M.}~\bibnamefont {G\"{a}chter}}, \bibinfo {author} {\bibfnamefont {J.}~\bibnamefont {Minguzzi}}, \bibinfo {author} {\bibfnamefont {S.}~\bibnamefont {Roschinski}}, \bibinfo {author} {\bibfnamefont {K.}~\bibnamefont {Sandholzer}}, \bibinfo {author} {\bibfnamefont {K.}~\bibnamefont {Viebahn}},\ and\ \bibinfo {author} {\bibfnamefont {T.}~\bibnamefont {Esslinger}},\ }\bibfield  {title} {\bibinfo {title} {Quantization and its breakdown in a hubbard–thouless pump},\ }\href {https://doi.org/10.1038/s41567-023-02145-w} {\bibfield  {journal} {\bibinfo  {journal} {Nature Physics}\ }\textbf {\bibinfo {volume} {19}},\ \bibinfo {pages} {1471–1475} (\bibinfo {year} {2023})}\BibitemShut {NoStop}%
\bibitem [{\citenamefont {Viebahn}\ \emph {et~al.}(2024)\citenamefont {Viebahn}, \citenamefont {Walter}, \citenamefont {Bertok}, \citenamefont {Zhu}, \citenamefont {G\"achter}, \citenamefont {Aligia}, \citenamefont {Heidrich-Meisner},\ and\ \citenamefont {Esslinger}}]{viebahntopopump}%
  \BibitemOpen
  \bibfield  {author} {\bibinfo {author} {\bibfnamefont {K.}~\bibnamefont {Viebahn}}, \bibinfo {author} {\bibfnamefont {A.-S.}\ \bibnamefont {Walter}}, \bibinfo {author} {\bibfnamefont {E.}~\bibnamefont {Bertok}}, \bibinfo {author} {\bibfnamefont {Z.}~\bibnamefont {Zhu}}, \bibinfo {author} {\bibfnamefont {M.}~\bibnamefont {G\"achter}}, \bibinfo {author} {\bibfnamefont {A.~A.}\ \bibnamefont {Aligia}}, \bibinfo {author} {\bibfnamefont {F.}~\bibnamefont {Heidrich-Meisner}},\ and\ \bibinfo {author} {\bibfnamefont {T.}~\bibnamefont {Esslinger}},\ }\bibfield  {title} {\bibinfo {title} {Interactions enable thouless pumping in a nonsliding lattice},\ }\href {https://doi.org/10.1103/PhysRevX.14.021049} {\bibfield  {journal} {\bibinfo  {journal} {Phys. Rev. X}\ }\textbf {\bibinfo {volume} {14}},\ \bibinfo {pages} {021049} (\bibinfo {year} {2024})}\BibitemShut {NoStop}%
\bibitem [{\citenamefont {Marra}\ and\ \citenamefont {Nitta}(2020)}]{marraTopologicallyQuantizedCurrent2020}%
  \BibitemOpen
  \bibfield  {author} {\bibinfo {author} {\bibfnamefont {P.}~\bibnamefont {Marra}}\ and\ \bibinfo {author} {\bibfnamefont {M.}~\bibnamefont {Nitta}},\ }\bibfield  {title} {\bibinfo {title} {Topologically quantized current in quasiperiodic {{Thouless}} pumps},\ }\href {https://doi.org/10.1103/PhysRevResearch.2.042035} {\bibfield  {journal} {\bibinfo  {journal} {Physical Review Research}\ }\textbf {\bibinfo {volume} {2}},\ \bibinfo {pages} {042035} (\bibinfo {year} {2020})}\BibitemShut {NoStop}%
\bibitem [{\citenamefont {Moustaj}\ \emph {et~al.}(2024)\citenamefont {Moustaj}, \citenamefont {Krebbekx},\ and\ \citenamefont {Smith}}]{moustaj2024anomalouspolarizationonedimensionalaperiodic}%
  \BibitemOpen
  \bibfield  {author} {\bibinfo {author} {\bibfnamefont {A.}~\bibnamefont {Moustaj}}, \bibinfo {author} {\bibfnamefont {J.~P.~J.}\ \bibnamefont {Krebbekx}},\ and\ \bibinfo {author} {\bibfnamefont {C.~M.}\ \bibnamefont {Smith}},\ }\href {https://arxiv.org/abs/2404.14916} {\bibinfo {title} {Anomalous polarization in one-dimensional aperiodic insulators}} (\bibinfo {year} {2024}),\ \Eprint {https://arxiv.org/abs/2404.14916} {arXiv:2404.14916 [cond-mat.dis-nn]} \BibitemShut {NoStop}%
\bibitem [{\citenamefont {Grabarits}\ \emph {et~al.}(2024)\citenamefont {Grabarits}, \citenamefont {Tak\'acs}, \citenamefont {Fulga},\ and\ \citenamefont {Asb\'oth}}]{grabarits2023floquetanderson}%
  \BibitemOpen
  \bibfield  {author} {\bibinfo {author} {\bibfnamefont {A.}~\bibnamefont {Grabarits}}, \bibinfo {author} {\bibfnamefont {A.}~\bibnamefont {Tak\'acs}}, \bibinfo {author} {\bibfnamefont {I.~C.}\ \bibnamefont {Fulga}},\ and\ \bibinfo {author} {\bibfnamefont {J.~K.}\ \bibnamefont {Asb\'oth}},\ }\bibfield  {title} {\bibinfo {title} {Floquet-anderson localization in the thouless pump and how to avoid it},\ }\href {https://doi.org/10.1103/PhysRevB.109.L180202} {\bibfield  {journal} {\bibinfo  {journal} {Phys. Rev. B}\ }\textbf {\bibinfo {volume} {109}},\ \bibinfo {pages} {L180202} (\bibinfo {year} {2024})}\BibitemShut {NoStop}%
\bibitem [{\citenamefont {Vuina}\ \emph {et~al.}(2024)\citenamefont {Vuina}, \citenamefont {Long}, \citenamefont {Crowley},\ and\ \citenamefont {Chandran}}]{vuina2024absence}%
  \BibitemOpen
  \bibfield  {author} {\bibinfo {author} {\bibfnamefont {D.}~\bibnamefont {Vuina}}, \bibinfo {author} {\bibfnamefont {D.~M.}\ \bibnamefont {Long}}, \bibinfo {author} {\bibfnamefont {P.~J.~D.}\ \bibnamefont {Crowley}},\ and\ \bibinfo {author} {\bibfnamefont {A.}~\bibnamefont {Chandran}},\ }\href@noop {} {\bibinfo {title} {Absence of disordered thouless pumps at finite frequency}} (\bibinfo {year} {2024}),\ \Eprint {https://arxiv.org/abs/2401.17395} {arXiv:2401.17395 [cond-mat.dis-nn]} \BibitemShut {NoStop}%
\bibitem [{\citenamefont {Avron}\ and\ \citenamefont {Elgart}(1999)}]{avron1999adiabatic}%
  \BibitemOpen
  \bibfield  {author} {\bibinfo {author} {\bibfnamefont {J.~E.}\ \bibnamefont {Avron}}\ and\ \bibinfo {author} {\bibfnamefont {A.}~\bibnamefont {Elgart}},\ }\bibfield  {title} {\bibinfo {title} {Adiabatic theorem without a gap condition},\ }\href@noop {} {\bibfield  {journal} {\bibinfo  {journal} {Communications in mathematical physics}\ }\textbf {\bibinfo {volume} {203}},\ \bibinfo {pages} {445} (\bibinfo {year} {1999})}\BibitemShut {NoStop}%
\bibitem [{\citenamefont {Avron}\ \emph {et~al.}(2012)\citenamefont {Avron}, \citenamefont {Fraas}, \citenamefont {Graf},\ and\ \citenamefont {Grech}}]{avron2012adiabatic}%
  \BibitemOpen
  \bibfield  {author} {\bibinfo {author} {\bibfnamefont {J.}~\bibnamefont {Avron}}, \bibinfo {author} {\bibfnamefont {M.}~\bibnamefont {Fraas}}, \bibinfo {author} {\bibfnamefont {G.}~\bibnamefont {Graf}},\ and\ \bibinfo {author} {\bibfnamefont {P.}~\bibnamefont {Grech}},\ }\bibfield  {title} {\bibinfo {title} {Adiabatic theorems for generators of contracting evolutions},\ }\href@noop {} {\bibfield  {journal} {\bibinfo  {journal} {Communications in mathematical physics}\ }\textbf {\bibinfo {volume} {314}},\ \bibinfo {pages} {163} (\bibinfo {year} {2012})}\BibitemShut {NoStop}%
\bibitem [{\citenamefont {Laughlin}(1981)}]{PhysRevB.23.5632}%
  \BibitemOpen
  \bibfield  {author} {\bibinfo {author} {\bibfnamefont {R.~B.}\ \bibnamefont {Laughlin}},\ }\bibfield  {title} {\bibinfo {title} {Quantized hall conductivity in two dimensions},\ }\href {https://doi.org/10.1103/PhysRevB.23.5632} {\bibfield  {journal} {\bibinfo  {journal} {Physical Review B}\ }\textbf {\bibinfo {volume} {23}},\ \bibinfo {pages} {5632} (\bibinfo {year} {1981})}\BibitemShut {NoStop}%
\bibitem [{\citenamefont {Kitaev}(2009)}]{10.1063/1.3149495}%
  \BibitemOpen
  \bibfield  {author} {\bibinfo {author} {\bibfnamefont {A.}~\bibnamefont {Kitaev}},\ }\bibfield  {title} {\bibinfo {title} {{Periodic table for topological insulators and superconductors}},\ }\href {https://doi.org/10.1063/1.3149495} {\bibfield  {journal} {\bibinfo  {journal} {AIP Conference Proceedings}\ }\textbf {\bibinfo {volume} {1134}},\ \bibinfo {pages} {22} (\bibinfo {year} {2009})},\ \Eprint {https://arxiv.org/abs/https://pubs.aip.org/aip/acp/article-pdf/1134/1/22/11584243/22\_1\_online.pdf} {https://pubs.aip.org/aip/acp/article-pdf/1134/1/22/11584243/22\_1\_online.pdf} \BibitemShut {NoStop}%
\bibitem [{\citenamefont {Scaffidi}\ \emph {et~al.}(2017)\citenamefont {Scaffidi}, \citenamefont {Parker},\ and\ \citenamefont {Vasseur}}]{PhysRevX.7.041048}%
  \BibitemOpen
  \bibfield  {author} {\bibinfo {author} {\bibfnamefont {T.}~\bibnamefont {Scaffidi}}, \bibinfo {author} {\bibfnamefont {D.~E.}\ \bibnamefont {Parker}},\ and\ \bibinfo {author} {\bibfnamefont {R.}~\bibnamefont {Vasseur}},\ }\bibfield  {title} {\bibinfo {title} {Gapless symmetry-protected topological order},\ }\href {https://doi.org/10.1103/PhysRevX.7.041048} {\bibfield  {journal} {\bibinfo  {journal} {Physical Review X}\ }\textbf {\bibinfo {volume} {7}},\ \bibinfo {pages} {041048} (\bibinfo {year} {2017})}\BibitemShut {NoStop}%
\bibitem [{\citenamefont {Kruthoff}\ \emph {et~al.}(2017)\citenamefont {Kruthoff}, \citenamefont {de~Boer}, \citenamefont {van Wezel}, \citenamefont {Kane},\ and\ \citenamefont {Slager}}]{Clas3}%
  \BibitemOpen
  \bibfield  {author} {\bibinfo {author} {\bibfnamefont {J.}~\bibnamefont {Kruthoff}}, \bibinfo {author} {\bibfnamefont {J.}~\bibnamefont {de~Boer}}, \bibinfo {author} {\bibfnamefont {J.}~\bibnamefont {van Wezel}}, \bibinfo {author} {\bibfnamefont {C.~L.}\ \bibnamefont {Kane}},\ and\ \bibinfo {author} {\bibfnamefont {R.-J.}\ \bibnamefont {Slager}},\ }\bibfield  {title} {\bibinfo {title} {Topological classification of crystalline insulators through band structure combinatorics},\ }\href {https://doi.org/10.1103/PhysRevX.7.041069} {\bibfield  {journal} {\bibinfo  {journal} {Physical Review X}\ }\textbf {\bibinfo {volume} {7}},\ \bibinfo {pages} {041069} (\bibinfo {year} {2017})}\BibitemShut {NoStop}%
\bibitem [{\citenamefont {Verresen}\ \emph {et~al.}(2021)\citenamefont {Verresen}, \citenamefont {Thorngren}, \citenamefont {Jones},\ and\ \citenamefont {Pollmann}}]{PhysRevX.11.041059}%
  \BibitemOpen
  \bibfield  {author} {\bibinfo {author} {\bibfnamefont {R.}~\bibnamefont {Verresen}}, \bibinfo {author} {\bibfnamefont {R.}~\bibnamefont {Thorngren}}, \bibinfo {author} {\bibfnamefont {N.~G.}\ \bibnamefont {Jones}},\ and\ \bibinfo {author} {\bibfnamefont {F.}~\bibnamefont {Pollmann}},\ }\bibfield  {title} {\bibinfo {title} {Gapless topological phases and symmetry-enriched quantum criticality},\ }\href {https://doi.org/10.1103/PhysRevX.11.041059} {\bibfield  {journal} {\bibinfo  {journal} {Physical Review X}\ }\textbf {\bibinfo {volume} {11}},\ \bibinfo {pages} {041059} (\bibinfo {year} {2021})}\BibitemShut {NoStop}%
\bibitem [{\citenamefont {Wauters}\ \emph {et~al.}(2019)\citenamefont {Wauters}, \citenamefont {Russomanno}, \citenamefont {Citro}, \citenamefont {Santoro},\ and\ \citenamefont {Privitera}}]{Wauters}%
  \BibitemOpen
  \bibfield  {author} {\bibinfo {author} {\bibfnamefont {M.~M.}\ \bibnamefont {Wauters}}, \bibinfo {author} {\bibfnamefont {A.}~\bibnamefont {Russomanno}}, \bibinfo {author} {\bibfnamefont {R.}~\bibnamefont {Citro}}, \bibinfo {author} {\bibfnamefont {G.~E.}\ \bibnamefont {Santoro}},\ and\ \bibinfo {author} {\bibfnamefont {L.}~\bibnamefont {Privitera}},\ }\bibfield  {title} {\bibinfo {title} {Localization, topology, and quantized transport in disordered floquet systems},\ }\href {https://doi.org/10.1103/PhysRevLett.123.266601} {\bibfield  {journal} {\bibinfo  {journal} {Physical Review Letters}\ }\textbf {\bibinfo {volume} {123}},\ \bibinfo {pages} {266601} (\bibinfo {year} {2019})}\BibitemShut {NoStop}%
\bibitem [{\citenamefont {Nakajima}\ \emph {et~al.}(2021)\citenamefont {Nakajima}, \citenamefont {Takei}, \citenamefont {Sakuma}, \citenamefont {Kuno}, \citenamefont {Marra},\ and\ \citenamefont {Takahashi}}]{nakajimaCompetitionInterplayTopology2021b}%
  \BibitemOpen
  \bibfield  {author} {\bibinfo {author} {\bibfnamefont {S.}~\bibnamefont {Nakajima}}, \bibinfo {author} {\bibfnamefont {N.}~\bibnamefont {Takei}}, \bibinfo {author} {\bibfnamefont {K.}~\bibnamefont {Sakuma}}, \bibinfo {author} {\bibfnamefont {Y.}~\bibnamefont {Kuno}}, \bibinfo {author} {\bibfnamefont {P.}~\bibnamefont {Marra}},\ and\ \bibinfo {author} {\bibfnamefont {Y.}~\bibnamefont {Takahashi}},\ }\bibfield  {title} {\bibinfo {title} {Competition and interplay between topology and quasi-periodic disorder in {{Thouless}} pumping of ultracold atoms},\ }\href {https://doi.org/10.1038/s41567-021-01229-9} {\bibfield  {journal} {\bibinfo  {journal} {Nature Physics}\ }\textbf {\bibinfo {volume} {17}},\ \bibinfo {pages} {844} (\bibinfo {year} {2021})}\BibitemShut {NoStop}%
\bibitem [{\citenamefont {Hayward}\ \emph {et~al.}(2021)\citenamefont {Hayward}, \citenamefont {Bertok}, \citenamefont {Schneider},\ and\ \citenamefont {{Heidrich-Meisner}}}]{haywardEffectDisorderTopological2021a}%
  \BibitemOpen
  \bibfield  {author} {\bibinfo {author} {\bibfnamefont {A.~L.~C.}\ \bibnamefont {Hayward}}, \bibinfo {author} {\bibfnamefont {E.}~\bibnamefont {Bertok}}, \bibinfo {author} {\bibfnamefont {U.}~\bibnamefont {Schneider}},\ and\ \bibinfo {author} {\bibfnamefont {F.}~\bibnamefont {{Heidrich-Meisner}}},\ }\bibfield  {title} {\bibinfo {title} {Effect of disorder on topological charge pumping in the {{Rice-Mele}} model},\ }\href {https://doi.org/10.1103/PhysRevA.103.043310} {\bibfield  {journal} {\bibinfo  {journal} {Physical Review A}\ }\textbf {\bibinfo {volume} {103}},\ \bibinfo {pages} {043310} (\bibinfo {year} {2021})}\BibitemShut {NoStop}%
\bibitem [{\citenamefont {Jitomirskaya}\ and\ \citenamefont {Last}(1998)}]{jitomirskaya1998anderson}%
  \BibitemOpen
  \bibfield  {author} {\bibinfo {author} {\bibfnamefont {S.~Y.}\ \bibnamefont {Jitomirskaya}}\ and\ \bibinfo {author} {\bibfnamefont {Y.}~\bibnamefont {Last}},\ }\bibfield  {title} {\bibinfo {title} {Anderson localization for the almost mathieu equation, iii. semi-uniform localization, continuity of gaps, and measure of the spectrum},\ }\href@noop {} {\bibfield  {journal} {\bibinfo  {journal} {Communications in mathematical physics}\ }\textbf {\bibinfo {volume} {195}},\ \bibinfo {pages} {1} (\bibinfo {year} {1998})}\BibitemShut {NoStop}%
\bibitem [{\citenamefont {Avila}\ and\ \citenamefont {Jitomirskaya}(2009)}]{avila2009ten}%
  \BibitemOpen
  \bibfield  {author} {\bibinfo {author} {\bibfnamefont {A.}~\bibnamefont {Avila}}\ and\ \bibinfo {author} {\bibfnamefont {S.}~\bibnamefont {Jitomirskaya}},\ }\bibfield  {title} {\bibinfo {title} {The ten martini problem},\ }\href@noop {} {\bibfield  {journal} {\bibinfo  {journal} {Annals of mathematics}\ ,\ \bibinfo {pages} {303}} (\bibinfo {year} {2009})}\BibitemShut {NoStop}%
\bibitem [{\citenamefont {Bellissard}\ \emph {et~al.}(1992)\citenamefont {Bellissard}, \citenamefont {Bovier},\ and\ \citenamefont {Jean-Michel}}]{bellissard92}%
  \BibitemOpen
  \bibfield  {author} {\bibinfo {author} {\bibfnamefont {J.}~\bibnamefont {Bellissard}}, \bibinfo {author} {\bibfnamefont {A.}~\bibnamefont {Bovier}},\ and\ \bibinfo {author} {\bibfnamefont {G.}~\bibnamefont {Jean-Michel}},\ }\bibfield  {title} {\bibinfo {title} {Gap labelling theorems for one dimensional discrete schrödinger operators},\ }\href {https://doi.org/10.1142/S0129055X92000029} {\bibfield  {journal} {\bibinfo  {journal} {Reviews in Mathematical Physics}\ }\textbf {\bibinfo {volume} {04}},\ \bibinfo {pages} {1} (\bibinfo {year} {1992})}\BibitemShut {NoStop}%
\bibitem [{\citenamefont {Avila}\ \emph {et~al.}(2023)\citenamefont {Avila}, \citenamefont {You},\ and\ \citenamefont {Zhou}}]{drytenmartininoncritical}%
  \BibitemOpen
  \bibfield  {author} {\bibinfo {author} {\bibfnamefont {A.}~\bibnamefont {Avila}}, \bibinfo {author} {\bibfnamefont {J.}~\bibnamefont {You}},\ and\ \bibinfo {author} {\bibfnamefont {Q.}~\bibnamefont {Zhou}},\ }\href {https://doi.org/10.48550/ARXIV.2306.16254} {\bibinfo {title} {Dry ten martini problem in the non-critical case}} (\bibinfo {year} {2023})\BibitemShut {NoStop}%
\bibitem [{\citenamefont {Han}(2018)}]{han2018dry}%
  \BibitemOpen
  \bibfield  {author} {\bibinfo {author} {\bibfnamefont {R.}~\bibnamefont {Han}},\ }\bibfield  {title} {\bibinfo {title} {Dry ten martini problem for the non-self-dual extended harper’s model},\ }\href@noop {} {\bibfield  {journal} {\bibinfo  {journal} {Transactions of the American Mathematical Society}\ }\textbf {\bibinfo {volume} {370}},\ \bibinfo {pages} {197} (\bibinfo {year} {2018})}\BibitemShut {NoStop}%
\bibitem [{\citenamefont {Roati}\ \emph {et~al.}(2008)\citenamefont {Roati}, \citenamefont {D’Errico}, \citenamefont {Fallani}, \citenamefont {Fattori}, \citenamefont {Fort}, \citenamefont {Zaccanti}, \citenamefont {Modugno}, \citenamefont {Modugno},\ and\ \citenamefont {Inguscio}}]{Roati2008}%
  \BibitemOpen
  \bibfield  {author} {\bibinfo {author} {\bibfnamefont {G.}~\bibnamefont {Roati}}, \bibinfo {author} {\bibfnamefont {C.}~\bibnamefont {D’Errico}}, \bibinfo {author} {\bibfnamefont {L.}~\bibnamefont {Fallani}}, \bibinfo {author} {\bibfnamefont {M.}~\bibnamefont {Fattori}}, \bibinfo {author} {\bibfnamefont {C.}~\bibnamefont {Fort}}, \bibinfo {author} {\bibfnamefont {M.}~\bibnamefont {Zaccanti}}, \bibinfo {author} {\bibfnamefont {G.}~\bibnamefont {Modugno}}, \bibinfo {author} {\bibfnamefont {M.}~\bibnamefont {Modugno}},\ and\ \bibinfo {author} {\bibfnamefont {M.}~\bibnamefont {Inguscio}},\ }\bibfield  {title} {\bibinfo {title} {Anderson localization of a non-interacting bose–einstein condensate},\ }\href {https://doi.org/10.1038/nature07071} {\bibfield  {journal} {\bibinfo  {journal} {Nature}\ }\textbf {\bibinfo {volume} {453}},\ \bibinfo {pages} {895–898} (\bibinfo {year} {2008})}\BibitemShut {NoStop}%
\bibitem [{\citenamefont {Schreiber}\ \emph {et~al.}(2015)\citenamefont {Schreiber}, \citenamefont {Hodgman}, \citenamefont {Bordia}, \citenamefont {Lüschen}, \citenamefont {Fischer}, \citenamefont {Vosk}, \citenamefont {Altman}, \citenamefont {Schneider},\ and\ \citenamefont {Bloch}}]{schreiber}%
  \BibitemOpen
  \bibfield  {author} {\bibinfo {author} {\bibfnamefont {M.}~\bibnamefont {Schreiber}}, \bibinfo {author} {\bibfnamefont {S.~S.}\ \bibnamefont {Hodgman}}, \bibinfo {author} {\bibfnamefont {P.}~\bibnamefont {Bordia}}, \bibinfo {author} {\bibfnamefont {H.~P.}\ \bibnamefont {Lüschen}}, \bibinfo {author} {\bibfnamefont {M.~H.}\ \bibnamefont {Fischer}}, \bibinfo {author} {\bibfnamefont {R.}~\bibnamefont {Vosk}}, \bibinfo {author} {\bibfnamefont {E.}~\bibnamefont {Altman}}, \bibinfo {author} {\bibfnamefont {U.}~\bibnamefont {Schneider}},\ and\ \bibinfo {author} {\bibfnamefont {I.}~\bibnamefont {Bloch}},\ }\bibfield  {title} {\bibinfo {title} {Observation of many-body localization of interacting fermions in a quasirandom optical lattice},\ }\href {https://doi.org/10.1126/science.aaa7432} {\bibfield  {journal} {\bibinfo  {journal} {Science}\ }\textbf {\bibinfo {volume} {349}},\ \bibinfo {pages} {842} (\bibinfo {year} {2015})}\BibitemShut {NoStop}%
\bibitem [{\citenamefont {Goblot}\ \emph {et~al.}(2020)\citenamefont {Goblot}, \citenamefont {Štrkalj}, \citenamefont {Pernet}, \citenamefont {Lado}, \citenamefont {Dorow}, \citenamefont {Lemaître}, \citenamefont {Le~Gratiet}, \citenamefont {Harouri}, \citenamefont {Sagnes}, \citenamefont {Ravets}, \citenamefont {Amo}, \citenamefont {Bloch},\ and\ \citenamefont {Zilberberg}}]{Goblot2020}%
  \BibitemOpen
  \bibfield  {author} {\bibinfo {author} {\bibfnamefont {V.}~\bibnamefont {Goblot}}, \bibinfo {author} {\bibfnamefont {A.}~\bibnamefont {Štrkalj}}, \bibinfo {author} {\bibfnamefont {N.}~\bibnamefont {Pernet}}, \bibinfo {author} {\bibfnamefont {J.~L.}\ \bibnamefont {Lado}}, \bibinfo {author} {\bibfnamefont {C.}~\bibnamefont {Dorow}}, \bibinfo {author} {\bibfnamefont {A.}~\bibnamefont {Lemaître}}, \bibinfo {author} {\bibfnamefont {L.}~\bibnamefont {Le~Gratiet}}, \bibinfo {author} {\bibfnamefont {A.}~\bibnamefont {Harouri}}, \bibinfo {author} {\bibfnamefont {I.}~\bibnamefont {Sagnes}}, \bibinfo {author} {\bibfnamefont {S.}~\bibnamefont {Ravets}}, \bibinfo {author} {\bibfnamefont {A.}~\bibnamefont {Amo}}, \bibinfo {author} {\bibfnamefont {J.}~\bibnamefont {Bloch}},\ and\ \bibinfo {author} {\bibfnamefont {O.}~\bibnamefont {Zilberberg}},\ }\bibfield  {title} {\bibinfo {title} {Emergence of criticality through a cascade of delocalization transitions in quasiperiodic chains},\ }\href
  {https://doi.org/10.1038/s41567-020-0908-7} {\bibfield  {journal} {\bibinfo  {journal} {Nature Physics}\ }\textbf {\bibinfo {volume} {16}},\ \bibinfo {pages} {832–836} (\bibinfo {year} {2020})}\BibitemShut {NoStop}%
\bibitem [{\citenamefont {Jitomirskaya}(1999)}]{jitomirskaya1999metal}%
  \BibitemOpen
  \bibfield  {author} {\bibinfo {author} {\bibfnamefont {S.~Y.}\ \bibnamefont {Jitomirskaya}},\ }\bibfield  {title} {\bibinfo {title} {Metal-insulator transition for the almost mathieu operator},\ }\href@noop {} {\bibfield  {journal} {\bibinfo  {journal} {Annals of Mathematics}\ ,\ \bibinfo {pages} {1159}} (\bibinfo {year} {1999})}\BibitemShut {NoStop}%
\bibitem [{\citenamefont {Aubry}\ and\ \citenamefont {Andr{\'e}}(1980)}]{aubry1980annals}%
  \BibitemOpen
  \bibfield  {author} {\bibinfo {author} {\bibfnamefont {S.}~\bibnamefont {Aubry}}\ and\ \bibinfo {author} {\bibfnamefont {G.}~\bibnamefont {Andr{\'e}}},\ }\bibfield  {title} {\bibinfo {title} {Analyticity breaking and anderson localization in incommensurate lattices},\ }\href@noop {} {\bibfield  {journal} {\bibinfo  {journal} {Ann. Israel Phys. Soc}\ }\textbf {\bibinfo {volume} {3}},\ \bibinfo {pages} {133} (\bibinfo {year} {1980})}\BibitemShut {NoStop}%
\bibitem [{\citenamefont {Zorzi}(2015)}]{zorziElementaryProofEquidistribution2015}%
  \BibitemOpen
  \bibfield  {author} {\bibinfo {author} {\bibfnamefont {A.}~\bibnamefont {Zorzi}},\ }\bibfield  {title} {\bibinfo {title} {An {{Elementary Proof}} for the {{Equidistribution Theorem}}},\ }\href {https://doi.org/10.1007/s00283-014-9505-x} {\bibfield  {journal} {\bibinfo  {journal} {Math Intelligencer}\ }\textbf {\bibinfo {volume} {37}},\ \bibinfo {pages} {1} (\bibinfo {year} {2015})}\BibitemShut {NoStop}%
\bibitem [{\citenamefont {Borgnia}\ \emph {et~al.}(2022)\citenamefont {Borgnia}, \citenamefont {Vishwanath},\ and\ \citenamefont {Slager}}]{paper1}%
  \BibitemOpen
  \bibfield  {author} {\bibinfo {author} {\bibfnamefont {D.~S.}\ \bibnamefont {Borgnia}}, \bibinfo {author} {\bibfnamefont {A.}~\bibnamefont {Vishwanath}},\ and\ \bibinfo {author} {\bibfnamefont {R.-J.}\ \bibnamefont {Slager}},\ }\bibfield  {title} {\bibinfo {title} {Rational approximations of quasiperiodicity via projected green's functions},\ }\href {https://doi.org/10.1103/PhysRevB.106.054204} {\bibfield  {journal} {\bibinfo  {journal} {Physical Review B}\ }\textbf {\bibinfo {volume} {106}},\ \bibinfo {pages} {054204} (\bibinfo {year} {2022})}\BibitemShut {NoStop}%
\bibitem [{\citenamefont {Borgnia}\ and\ \citenamefont {Slager}(2023)}]{paper2}%
  \BibitemOpen
  \bibfield  {author} {\bibinfo {author} {\bibfnamefont {D.~S.}\ \bibnamefont {Borgnia}}\ and\ \bibinfo {author} {\bibfnamefont {R.-J.}\ \bibnamefont {Slager}},\ }\bibfield  {title} {\bibinfo {title} {Localization as a consequence of quasiperiodic bulk-bulk correspondence},\ }\href {https://doi.org/10.1103/PhysRevB.107.085111} {\bibfield  {journal} {\bibinfo  {journal} {Physical Review B}\ }\textbf {\bibinfo {volume} {107}},\ \bibinfo {pages} {085111} (\bibinfo {year} {2023})}\BibitemShut {NoStop}%
\bibitem [{\citenamefont {Kato}(1950)}]{Kato1950}%
  \BibitemOpen
  \bibfield  {author} {\bibinfo {author} {\bibfnamefont {T.}~\bibnamefont {Kato}},\ }\bibfield  {title} {\bibinfo {title} {On the adiabatic theorem of quantum mechanics},\ }\href {https://doi.org/10.1143/JPSJ.5.435} {\bibfield  {journal} {\bibinfo  {journal} {Journal of the Physical Society of Japan}\ }\textbf {\bibinfo {volume} {5}},\ \bibinfo {pages} {435} (\bibinfo {year} {1950})}\BibitemShut {NoStop}%
\bibitem [{\citenamefont {Hagedorn}(1991)}]{hagedorn1991proof}%
  \BibitemOpen
  \bibfield  {author} {\bibinfo {author} {\bibfnamefont {G.~A.}\ \bibnamefont {Hagedorn}},\ }\bibfield  {title} {\bibinfo {title} {Proof of the landau-zener formula in an adiabatic limit with small eigenvalue gaps},\ }\href@noop {} {\bibfield  {journal} {\bibinfo  {journal} {Communications in mathematical physics}\ }\textbf {\bibinfo {volume} {136}},\ \bibinfo {pages} {433} (\bibinfo {year} {1991})}\BibitemShut {NoStop}%
\bibitem [{\citenamefont {Anderson}(1958)}]{PhysRev.109.1492}%
  \BibitemOpen
  \bibfield  {author} {\bibinfo {author} {\bibfnamefont {P.~W.}\ \bibnamefont {Anderson}},\ }\bibfield  {title} {\bibinfo {title} {Absence of diffusion in certain random lattices},\ }\href {https://doi.org/10.1103/PhysRev.109.1492} {\bibfield  {journal} {\bibinfo  {journal} {Physical Review}\ }\textbf {\bibinfo {volume} {109}},\ \bibinfo {pages} {1492} (\bibinfo {year} {1958})}\BibitemShut {NoStop}%
\bibitem [{\citenamefont {Avila}\ \emph {et~al.}(2022)\citenamefont {Avila}, \citenamefont {Damanik},\ and\ \citenamefont {Gorodetski}}]{avilanewresult}%
  \BibitemOpen
  \bibfield  {author} {\bibinfo {author} {\bibfnamefont {A.}~\bibnamefont {Avila}}, \bibinfo {author} {\bibfnamefont {D.}~\bibnamefont {Damanik}},\ and\ \bibinfo {author} {\bibfnamefont {A.}~\bibnamefont {Gorodetski}},\ }\href {https://doi.org/10.48550/ARXIV.2211.02173} {\bibinfo {title} {The spectrum of schrödinger operators with randomly perturbed ergodic potentials}} (\bibinfo {year} {2022})\BibitemShut {NoStop}%
\bibitem [{\citenamefont {Shapiro}(2020)}]{articleshapiro}%
  \BibitemOpen
  \bibfield  {author} {\bibinfo {author} {\bibfnamefont {J.}~\bibnamefont {Shapiro}},\ }\bibfield  {title} {\bibinfo {title} {The topology of mobility-gapped insulators},\ }\href {https://doi.org/10.1007/s11005-020-01314-9} {\bibfield  {journal} {\bibinfo  {journal} {Letters in Mathematical Physics}\ }\textbf {\bibinfo {volume} {110}} (\bibinfo {year} {2020})}\BibitemShut {NoStop}%
\bibitem [{\citenamefont {Bianco}\ and\ \citenamefont {Resta}(2011)}]{PhysRevB.84.241106}%
  \BibitemOpen
  \bibfield  {author} {\bibinfo {author} {\bibfnamefont {R.}~\bibnamefont {Bianco}}\ and\ \bibinfo {author} {\bibfnamefont {R.}~\bibnamefont {Resta}},\ }\bibfield  {title} {\bibinfo {title} {Mapping topological order in coordinate space},\ }\href {https://doi.org/10.1103/PhysRevB.84.241106} {\bibfield  {journal} {\bibinfo  {journal} {Phys. Rev. B}\ }\textbf {\bibinfo {volume} {84}},\ \bibinfo {pages} {241106} (\bibinfo {year} {2011})}\BibitemShut {NoStop}%
\bibitem [{\citenamefont {Bordia}\ \emph {et~al.}(2016)\citenamefont {Bordia}, \citenamefont {L\"uschen}, \citenamefont {Hodgman}, \citenamefont {Schreiber}, \citenamefont {Bloch},\ and\ \citenamefont {Schneider}}]{Bordia2016}%
  \BibitemOpen
  \bibfield  {author} {\bibinfo {author} {\bibfnamefont {P.}~\bibnamefont {Bordia}}, \bibinfo {author} {\bibfnamefont {H.~P.}\ \bibnamefont {L\"uschen}}, \bibinfo {author} {\bibfnamefont {S.~S.}\ \bibnamefont {Hodgman}}, \bibinfo {author} {\bibfnamefont {M.}~\bibnamefont {Schreiber}}, \bibinfo {author} {\bibfnamefont {I.}~\bibnamefont {Bloch}},\ and\ \bibinfo {author} {\bibfnamefont {U.}~\bibnamefont {Schneider}},\ }\bibfield  {title} {\bibinfo {title} {Coupling identical one-dimensional many-body localized systems},\ }\href {https://doi.org/10.1103/PhysRevLett.116.140401} {\bibfield  {journal} {\bibinfo  {journal} {Phys. Rev. Lett.}\ }\textbf {\bibinfo {volume} {116}},\ \bibinfo {pages} {140401} (\bibinfo {year} {2016})}\BibitemShut {NoStop}%
\bibitem [{\citenamefont {Prodan}(2015{\natexlab{b}})}]{prodanVirtual2015}%
  \BibitemOpen
  \bibfield  {author} {\bibinfo {author} {\bibfnamefont {E.}~\bibnamefont {Prodan}},\ }\bibfield  {title} {\bibinfo {title} {Virtual topological insulators with real quantized physics},\ }\href {https://doi.org/10.1103/PhysRevB.91.245104} {\bibfield  {journal} {\bibinfo  {journal} {Phys. Rev. B}\ }\textbf {\bibinfo {volume} {91}},\ \bibinfo {pages} {245104} (\bibinfo {year} {2015}{\natexlab{b}})}\BibitemShut {NoStop}%
\bibitem [{\citenamefont {Magnus}(1954)}]{https://doi.org/10.1002/cpa.3160070404}%
  \BibitemOpen
  \bibfield  {author} {\bibinfo {author} {\bibfnamefont {W.}~\bibnamefont {Magnus}},\ }\bibfield  {title} {\bibinfo {title} {On the exponential solution of differential equations for a linear operator},\ }\href {https://doi.org/https://doi.org/10.1002/cpa.3160070404} {\bibfield  {journal} {\bibinfo  {journal} {Communications on Pure and Applied Mathematics}\ }\textbf {\bibinfo {volume} {7}},\ \bibinfo {pages} {649} (\bibinfo {year} {1954})}\BibitemShut {NoStop}%
\bibitem [{\citenamefont {Wannier}(1960)}]{Wannier1960432}%
  \BibitemOpen
  \bibfield  {author} {\bibinfo {author} {\bibfnamefont {G.~H.}\ \bibnamefont {Wannier}},\ }\bibfield  {title} {\bibinfo {title} {Wave functions and effective hamiltonian for bloch electrons in an electric field},\ }\href {https://doi.org/10.1103/PhysRev.117.432} {\bibfield  {journal} {\bibinfo  {journal} {Physical Review}\ }\textbf {\bibinfo {volume} {117}},\ \bibinfo {pages} {432 – 439} (\bibinfo {year} {1960})},\ \bibinfo {note} {cited by: 696}\BibitemShut {NoStop}%
\bibitem [{\citenamefont {Gottlob}\ and\ \citenamefont {Schneider}(2023)}]{gottlobPRB}%
  \BibitemOpen
  \bibfield  {author} {\bibinfo {author} {\bibfnamefont {E.}~\bibnamefont {Gottlob}}\ and\ \bibinfo {author} {\bibfnamefont {U.}~\bibnamefont {Schneider}},\ }\bibfield  {title} {\bibinfo {title} {Hubbard models for quasicrystalline potentials},\ }\href {https://doi.org/10.1103/PhysRevB.107.144202} {\bibfield  {journal} {\bibinfo  {journal} {Physical Review B}\ }\textbf {\bibinfo {volume} {107}},\ \bibinfo {pages} {144202} (\bibinfo {year} {2023})}\BibitemShut {NoStop}%
\bibitem [{\citenamefont {Sbroscia}\ \emph {et~al.}(2020)\citenamefont {Sbroscia}, \citenamefont {Viebahn}, \citenamefont {Carter}, \citenamefont {Yu}, \citenamefont {Gaunt},\ and\ \citenamefont {Schneider}}]{sbrosciaQC}%
  \BibitemOpen
  \bibfield  {author} {\bibinfo {author} {\bibfnamefont {M.}~\bibnamefont {Sbroscia}}, \bibinfo {author} {\bibfnamefont {K.}~\bibnamefont {Viebahn}}, \bibinfo {author} {\bibfnamefont {E.}~\bibnamefont {Carter}}, \bibinfo {author} {\bibfnamefont {J.-C.}\ \bibnamefont {Yu}}, \bibinfo {author} {\bibfnamefont {A.}~\bibnamefont {Gaunt}},\ and\ \bibinfo {author} {\bibfnamefont {U.}~\bibnamefont {Schneider}},\ }\bibfield  {title} {\bibinfo {title} {Observing localization in a 2d quasicrystalline optical lattice},\ }\href {https://doi.org/10.1103/PhysRevLett.125.200604} {\bibfield  {journal} {\bibinfo  {journal} {Physical Review Letters}\ }\textbf {\bibinfo {volume} {125}},\ \bibinfo {pages} {200604} (\bibinfo {year} {2020})}\BibitemShut {NoStop}%
\bibitem [{\citenamefont {Viebahn}\ \emph {et~al.}(2019)\citenamefont {Viebahn}, \citenamefont {Sbroscia}, \citenamefont {Carter}, \citenamefont {Yu},\ and\ \citenamefont {Schneider}}]{ViehbanQC}%
  \BibitemOpen
  \bibfield  {author} {\bibinfo {author} {\bibfnamefont {K.}~\bibnamefont {Viebahn}}, \bibinfo {author} {\bibfnamefont {M.}~\bibnamefont {Sbroscia}}, \bibinfo {author} {\bibfnamefont {E.}~\bibnamefont {Carter}}, \bibinfo {author} {\bibfnamefont {J.-C.}\ \bibnamefont {Yu}},\ and\ \bibinfo {author} {\bibfnamefont {U.}~\bibnamefont {Schneider}},\ }\bibfield  {title} {\bibinfo {title} {Matter-wave diffraction from a quasicrystalline optical lattice},\ }\href {https://doi.org/10.1103/PhysRevLett.122.110404} {\bibfield  {journal} {\bibinfo  {journal} {Physical Review Letters}\ }\textbf {\bibinfo {volume} {122}},\ \bibinfo {pages} {110404} (\bibinfo {year} {2019})}\BibitemShut {NoStop}%
\bibitem [{\citenamefont {Navon}\ \emph {et~al.}(2021)\citenamefont {Navon}, \citenamefont {Smith},\ and\ \citenamefont {Hadzibabic}}]{Navon2021}%
  \BibitemOpen
  \bibfield  {author} {\bibinfo {author} {\bibfnamefont {N.}~\bibnamefont {Navon}}, \bibinfo {author} {\bibfnamefont {R.~P.}\ \bibnamefont {Smith}},\ and\ \bibinfo {author} {\bibfnamefont {Z.}~\bibnamefont {Hadzibabic}},\ }\bibfield  {title} {\bibinfo {title} {Quantum gases in optical boxes},\ }\href {https://doi.org/10.1038/s41567-021-01403-z} {\bibfield  {journal} {\bibinfo  {journal} {Nature Physics}\ }\textbf {\bibinfo {volume} {17}},\ \bibinfo {pages} {1334–1341} (\bibinfo {year} {2021})}\BibitemShut {NoStop}%
\bibitem [{\citenamefont {Chin}\ \emph {et~al.}(2010)\citenamefont {Chin}, \citenamefont {Grimm}, \citenamefont {Julienne},\ and\ \citenamefont {Tiesinga}}]{ChinFeshbach}%
  \BibitemOpen
  \bibfield  {author} {\bibinfo {author} {\bibfnamefont {C.}~\bibnamefont {Chin}}, \bibinfo {author} {\bibfnamefont {R.}~\bibnamefont {Grimm}}, \bibinfo {author} {\bibfnamefont {P.}~\bibnamefont {Julienne}},\ and\ \bibinfo {author} {\bibfnamefont {E.}~\bibnamefont {Tiesinga}},\ }\bibfield  {title} {\bibinfo {title} {Feshbach resonances in ultracold gases},\ }\href {https://doi.org/10.1103/RevModPhys.82.1225} {\bibfield  {journal} {\bibinfo  {journal} {Rev. Mod. Phys.}\ }\textbf {\bibinfo {volume} {82}},\ \bibinfo {pages} {1225} (\bibinfo {year} {2010})}\BibitemShut {NoStop}%
\bibitem [{\citenamefont {Hofstadter}(1976)}]{PhysRevB.14.2239}%
  \BibitemOpen
  \bibfield  {author} {\bibinfo {author} {\bibfnamefont {D.~R.}\ \bibnamefont {Hofstadter}},\ }\bibfield  {title} {\bibinfo {title} {Energy levels and wave functions of bloch electrons in rational and irrational magnetic fields},\ }\href {https://doi.org/10.1103/PhysRevB.14.2239} {\bibfield  {journal} {\bibinfo  {journal} {Phys. Rev. B}\ }\textbf {\bibinfo {volume} {14}},\ \bibinfo {pages} {2239} (\bibinfo {year} {1976})}\BibitemShut {NoStop}%
\bibitem [{\citenamefont {Ryu}\ \emph {et~al.}(2010)\citenamefont {Ryu}, \citenamefont {Schnyder}, \citenamefont {Furusaki},\ and\ \citenamefont {Ludwig}}]{clas1b}%
  \BibitemOpen
  \bibfield  {author} {\bibinfo {author} {\bibfnamefont {S.}~\bibnamefont {Ryu}}, \bibinfo {author} {\bibfnamefont {A.~P.}\ \bibnamefont {Schnyder}}, \bibinfo {author} {\bibfnamefont {A.}~\bibnamefont {Furusaki}},\ and\ \bibinfo {author} {\bibfnamefont {A.~W.}\ \bibnamefont {Ludwig}},\ }\bibfield  {title} {\bibinfo {title} {Topological insulators and superconductors: tenfold way and dimensional hierarchy},\ }\href@noop {} {\bibfield  {journal} {\bibinfo  {journal} {New Journal of Physics}\ }\textbf {\bibinfo {volume} {12}},\ \bibinfo {pages} {065010} (\bibinfo {year} {2010})}\BibitemShut {NoStop}%
\bibitem [{\citenamefont {An}\ \emph {et~al.}(2018)\citenamefont {An}, \citenamefont {Meier}, \citenamefont {Ang'ong'a},\ and\ \citenamefont {Gadway}}]{PhysRevLett.120.040407}%
  \BibitemOpen
  \bibfield  {author} {\bibinfo {author} {\bibfnamefont {F.~A.}\ \bibnamefont {An}}, \bibinfo {author} {\bibfnamefont {E.~J.}\ \bibnamefont {Meier}}, \bibinfo {author} {\bibfnamefont {J.}~\bibnamefont {Ang'ong'a}},\ and\ \bibinfo {author} {\bibfnamefont {B.}~\bibnamefont {Gadway}},\ }\bibfield  {title} {\bibinfo {title} {Correlated dynamics in a synthetic lattice of momentum states},\ }\href {https://doi.org/10.1103/PhysRevLett.120.040407} {\bibfield  {journal} {\bibinfo  {journal} {Phys. Rev. Lett.}\ }\textbf {\bibinfo {volume} {120}},\ \bibinfo {pages} {040407} (\bibinfo {year} {2018})}\BibitemShut {NoStop}%
\bibitem [{\citenamefont {Liu}\ \emph {et~al.}(2013)\citenamefont {Liu}, \citenamefont {Bergholtz},\ and\ \citenamefont {Kapit}}]{PhysRevB.88.205101}%
  \BibitemOpen
  \bibfield  {author} {\bibinfo {author} {\bibfnamefont {Z.}~\bibnamefont {Liu}}, \bibinfo {author} {\bibfnamefont {E.~J.}\ \bibnamefont {Bergholtz}},\ and\ \bibinfo {author} {\bibfnamefont {E.}~\bibnamefont {Kapit}},\ }\bibfield  {title} {\bibinfo {title} {Non-abelian fractional chern insulators from long-range interactions},\ }\href {https://doi.org/10.1103/PhysRevB.88.205101} {\bibfield  {journal} {\bibinfo  {journal} {Phys. Rev. B}\ }\textbf {\bibinfo {volume} {88}},\ \bibinfo {pages} {205101} (\bibinfo {year} {2013})}\BibitemShut {NoStop}%
\bibitem [{\citenamefont {Kapit}\ and\ \citenamefont {Mueller}(2010)}]{PhysRevLett.105.215303}%
  \BibitemOpen
  \bibfield  {author} {\bibinfo {author} {\bibfnamefont {E.}~\bibnamefont {Kapit}}\ and\ \bibinfo {author} {\bibfnamefont {E.}~\bibnamefont {Mueller}},\ }\bibfield  {title} {\bibinfo {title} {Exact parent hamiltonian for the quantum hall states in a lattice},\ }\href {https://doi.org/10.1103/PhysRevLett.105.215303} {\bibfield  {journal} {\bibinfo  {journal} {Phys. Rev. Lett.}\ }\textbf {\bibinfo {volume} {105}},\ \bibinfo {pages} {215303} (\bibinfo {year} {2010})}\BibitemShut {NoStop}%
\bibitem [{\citenamefont {Grushin}\ \emph {et~al.}(2012)\citenamefont {Grushin}, \citenamefont {Neupert}, \citenamefont {Chamon},\ and\ \citenamefont {Mudry}}]{PhysRevB.86.205125}%
  \BibitemOpen
  \bibfield  {author} {\bibinfo {author} {\bibfnamefont {A.~G.}\ \bibnamefont {Grushin}}, \bibinfo {author} {\bibfnamefont {T.}~\bibnamefont {Neupert}}, \bibinfo {author} {\bibfnamefont {C.}~\bibnamefont {Chamon}},\ and\ \bibinfo {author} {\bibfnamefont {C.}~\bibnamefont {Mudry}},\ }\bibfield  {title} {\bibinfo {title} {Enhancing the stability of a fractional chern insulator against competing phases},\ }\href {https://doi.org/10.1103/PhysRevB.86.205125} {\bibfield  {journal} {\bibinfo  {journal} {Phys. Rev. B}\ }\textbf {\bibinfo {volume} {86}},\ \bibinfo {pages} {205125} (\bibinfo {year} {2012})}\BibitemShut {NoStop}%
\bibitem [{\citenamefont {Liu}\ \emph {et~al.}(2012)\citenamefont {Liu}, \citenamefont {Bergholtz}, \citenamefont {Fan},\ and\ \citenamefont {L\"auchli}}]{PhysRevLett.109.186805}%
  \BibitemOpen
  \bibfield  {author} {\bibinfo {author} {\bibfnamefont {Z.}~\bibnamefont {Liu}}, \bibinfo {author} {\bibfnamefont {E.~J.}\ \bibnamefont {Bergholtz}}, \bibinfo {author} {\bibfnamefont {H.}~\bibnamefont {Fan}},\ and\ \bibinfo {author} {\bibfnamefont {A.~M.}\ \bibnamefont {L\"auchli}},\ }\bibfield  {title} {\bibinfo {title} {Fractional chern insulators in topological flat bands with higher chern number},\ }\href {https://doi.org/10.1103/PhysRevLett.109.186805} {\bibfield  {journal} {\bibinfo  {journal} {Phys. Rev. Lett.}\ }\textbf {\bibinfo {volume} {109}},\ \bibinfo {pages} {186805} (\bibinfo {year} {2012})}\BibitemShut {NoStop}%
\bibitem [{\citenamefont {Gonçalves}\ \emph {et~al.}(2022)\citenamefont {Gonçalves}, \citenamefont {Amorim}, \citenamefont {Castro},\ and\ \citenamefont {Ribeiro}}]{goncalves2021hidden}%
  \BibitemOpen
  \bibfield  {author} {\bibinfo {author} {\bibfnamefont {M.}~\bibnamefont {Gonçalves}}, \bibinfo {author} {\bibfnamefont {B.}~\bibnamefont {Amorim}}, \bibinfo {author} {\bibfnamefont {E.~V.}\ \bibnamefont {Castro}},\ and\ \bibinfo {author} {\bibfnamefont {P.}~\bibnamefont {Ribeiro}},\ }\bibfield  {title} {\bibinfo {title} {{Hidden dualities in 1D quasiperiodic lattice models}},\ }\href {https://doi.org/10.21468/SciPostPhys.13.3.046} {\bibfield  {journal} {\bibinfo  {journal} {SciPost Physics}\ }\textbf {\bibinfo {volume} {13}},\ \bibinfo {pages} {046} (\bibinfo {year} {2022})}\BibitemShut {NoStop}%
\end{thebibliography}%

\appendix

\section{Sign of gap labels from perturbation theory} \label{app:gaplabels}
We showed in \cref{section:configspace} how resonant pairs of lattice sites coupled by the effective $n$-th order tunneling element $J_{eff}^{(n)}$ give rise to pairs of energy gaps in the spectrum of the AAH chain at energies $ \pm 2V \cos{(n\pi\beta)} $. According to the gap labeling theorem, each of these gaps is characterised by a Chern number, whose sign dictates the direction of the current arising under adiabatic pumping when all states below that gap are occupied.

\begin{figure}
    \centering
    \includegraphics[width = \linewidth]{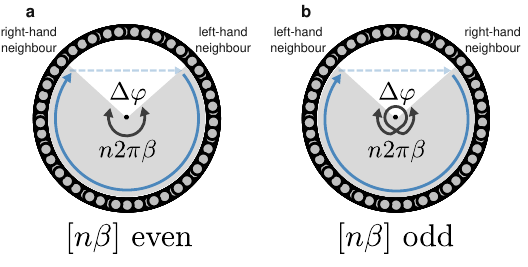}
    \caption{\textbf{Dependence of the signs of the gap labels on the parity of $[n\beta]$}: 
    In the circle forming configuration space, the Landau-Zener transition for states below the gap always goes from left to right. However, depending on $[n\beta]$, the position of left- and right-hand neighbours on the circle can switch. Figures \textbf{a} and \textbf{a} show two different gaps at similar energies to illustrate the two cases.(\textbf{a}) When $[n\beta]$ is even, the associated Landau-Zener transition involves transport from the right-hand resonant neighbor to the left-hand one. The resulting quantized current flows in the negative direction and is therefore associated with a negative Chern number. (\textbf{b}) When $[n\beta]$ is odd, the situation is reversed, and the quantized current flows in the positive direction such that the associated Chern number is positive. Note that in contrast to \cref{fig:fig_pumping}, we are here considering states below the gap. }
    \label{fig:labels}
\end{figure}

In the case $[n\beta]$ even, let us for instance fill up all states up to the energy  $-2V \cos{(n\pi \beta)}$, see \cref{fig:labels}, which for the case discussed in \cref{fig:fig_pumping} ($n=1$, $\beta>0.5$) is positive. For states below this gap, the pumping induces Landau-Zener hoppings in the negative direction, where each Landau-Zener transition transfers the states from the right-hand resonant neighbor to the left-hand one, see \cref{fig:labels}a. In this case we can ignore all other gaps, as they do not influence the overall current since their adjacent bands are either both filled or both empty. The pumping thus results in a quantized current in the negative direction and this gap is therefore associated with a negative Chern number. The same reasoning can be applied to the state filled up to $2V \cos{(n\pi \beta)}$, leading to a positive gap label.

In the case $[n\beta]$ odd, the resonant neighbors are flipped, see \cref{fig:labels}b, and when filling all the states up to $-2V \cos{(n\pi \beta)}$, the adiabatic Landau-Zener transition transfers the states from the left- to the right-hand neighbor. In this case, the quantized current flows in the positive direction and is therefore associated to a positive Chern number.

Therefore, for the sign of the gap labels to match the direction of the quantized current, one has to treat the even and odd cases of $[n\beta]$ separately, as done in \cref{eq:gaps}.

\section{Estimation of critical disorder strengths for first-order $C=+1$ bands} \label{appendix:disorder}

We use the configuration space pictures to estimate the critical disorder strength for the the $2$ first-order bands with $C=+1$, defined by the $n=\pm1$ gaps, in the case $\beta>1/2$. Note, the cases $\beta<1/2$ can be obtained by sending $\beta\rightarrow 1- \beta$ and inverting the sign of all Chern numbers.

 Let us consider $3$ consecutive sites with indices $i=-1, 0, 1$, each being offset by an onsite disorder of strength $w_i$. The phase $\varphi_{-1,0}$, where  the $-1$ and $0$ sites  are resonant, is given by 
\begin{align}
	2V\cos{(\varphi_{-1,0})} +w_{-1} = 2V\cos{(\varphi_{-1,0} + 2\pi \beta)} + w_0   \,,
\end{align}
and the corresponding condition for sites $0$ and $1$ is 
\begin{align}
	2V\cos{(\varphi_{0,1})} +w_0 = 2V\cos{(\varphi_{0,1} + 2\pi \beta)} + w_1 . 
\end{align}
The shifted  resonance conditions $\varphi_{-1,0}$ and $\varphi_{0,1}$  can thus be written as:
\begin{align} \label{eq:disorder_resonances_top}
	\varphi_{-1,0} = \pi (1-\beta) + \arcsin{\left(\frac{w_0-w_{-1}}{4V \sin{\left(\pi \beta\right)}}\right)}  \\
	\varphi_{0,1} = -\pi (1-\beta) + \arcsin{\left(\frac{w_1-w_0}{4V \sin{\left(\pi \beta\right)}}\right)} 
\end{align}

for the top $C=+1$ band and 

\begin{align} \label{eq:disorder_resonances_bottom}
	\varphi_{-1,0} = \pi \beta + \arcsin{\left(\frac{w_0-w_{-1}}{4V \sin{\left(\pi \beta\right)}}\right)}  \\
	\varphi_{0,1} = -\pi \beta + \arcsin{\left(\frac{w_1-w_0}{4V \sin{\left(\pi \beta\right)}}\right)} 
\end{align}
for the bottom $C=+1$ band.

As discussed in \cref{section:disorder}, the breakdown of quantized transport occurs once $\Delta \varphi = \varphi_{-1,0}- \varphi_{0,1} =0$. For bounded local disorder this becomes possible as soon as the disorder is strong enough to shift the individual resonances to $\varphi_{-1,0}= \varphi_{0,1} = 0$. This condition leads to the following estimate for the critical disorder strength for the top and bottom $C=+1$ bands:
\begin{equation}\label{eq:critical_c_1}
	W_{crit}^{C = +1} = 2V\sin^2{\left(\pi \beta\right)}.
\end{equation}

\end{document}